\documentclass[structabstract]{aa}  
\usepackage{graphicx}
\usepackage{natbib}
\bibpunct{(}{)}{;}{a}{}{,} 
\usepackage{amsmath}
\usepackage{txfonts}
\begin{document}
   \title{The standard model of low-mass star formation applied to massive stars: a multi-wavelength picture of AFGL\,2591}

   \author{K. G. Johnston \inst{1},
          D. S. Shepherd \inst{2},
          T. P. Robitaille \inst{1}
           \and
          K. Wood \inst{3}
          }

   \institute{Max Planck Institute for Astronomy, K\"onigstuhl 17, D-69117 Heidelberg, Germany \\
              \email{johnston@mpia.de}
         \and
             National Radio Astronomy Observatory, 1003 Lopezville Rd, Socorro, New Mexico 87801, USA
         \and
             School of Physics \& Astronomy, University of St Andrews, North Haugh, St Andrews, KY16 9SS, UK \\    
           }

   \date{Accepted 27 Nov 2012}

  \abstract{While it is currently unclear from a theoretical standpoint which forces and processes dominate the formation of high-mass stars, and hence determine the mode in which they form, much of the recent observational evidence suggests that massive stars are born in a similar manner to their low-mass counterparts.}
  {This paper aims to investigate the hypothesis that the embedded luminous star AFGL\,2591-VLA\,3 (2.3$\times10^{5}$\,L$_{\odot}$ at 3.33\,kpc) is forming according to a scaled-up version of a low-mass star formation scenario.}
  {We present multi-configuration Very Large Array 3.6\,cm and 7\,mm, as well as Combined Array for Research in Millimeter Astronomy C$^{18}$O and 3\,mm continuum observations to investigate the morphology and kinematics of the ionized gas, dust, and molecular gas around AFGL\,2591. We also compare our results to ancillary Gemini North near-IR images, and model the near-IR to sub-mm Spectral Energy distribution (SED) and Two Micron All Sky Survey (2MASS) image profiles of AFGL\,2591 using a Monte-Carlo dust continuum radiative transfer code. }
  {The observed 3.6 cm images uncover for the first time that the central powering source AFGL\,2591-VLA\,3 has a compact core plus collimated jet morphology, extending 4000 AU eastward from the central source with an opening angle of $<10^{\circ}$ at this radius. However, at 7\,mm VLA\,3 does not show a jet morphology, but instead compact ($< 500$\,AU) emission, some of which ($<$0.57\,mJy of 2.9\,mJy) is estimated to be from dust emission. The spectral index of AFGL\,2591-VLA\,3 between 3.6\,cm and 7\,mm was found to be between 0.4 and 0.5, similar to that of an ionized wind. If the 3.6\,cm emission is modelled as an ionized jet, the jet has almost enough momentum to drive the larger-scale flow. However, assuming a shock efficiency of 10\%, the momentum rate of the jet is not sufficient to ionize itself via only shocks, and thus a significant portion of the emission is instead likely created in a photoionized wind. The C$^{18}$O emission uncovers dense entrained material in the outflow(s) from these young stars. The main features of the SED and 2MASS images of AFGL\,2591-VLA\,3 are also reproduced by our model dust geometry of a rotationally flattened envelope with and without a disk.}
 {The above results are consistent with a picture of massive star formation similar to that seen for low-mass protostars. However, within its envelope, AFGL\,2591-VLA\,3 contains at least four other young stars, constituting a small cluster. Therefore  it appears that AFGL\,2591-VLA\,3 may be able to source its accreting material from a shared gas reservoir while still exhibiting the phenomena expected during the formation of low-mass stars.}
    
 
\titlerunning{A multi-wavelength picture of AFGL 2591}
\authorrunning{K. G. Johnston et al.}
 
\maketitle
 
\section{Introduction \label{intro}}

Does the formation of a massive star differ significantly from that of a low-mass star? Arguably, all studies of high mass ($\ge$8\,M$_{\sun}$) star formation are centred upon this question. There are several possible reasons to expect differences at higher masses, one being that a massive star is thought to continue to accrete material after reaching the zero-age main sequence \citep[ZAMS, e.g.][]{hosokawa10}, which is a consequence of its Kelvin-Helmholtz contraction time-scale being shorter than its accretion time-scale. Therefore, several processes such as radiation pressure \citep[e.g.][]{yorke02} and ionization \citep[e.g.][]{keto02} may halt, decrease or alter accretion on to the star. However, in the earlier stages of protostellar evolution, \citet{hosokawa10} also find that the central accreting stars in their simulations become bloated due to accretion, so that the effective temperature and UV luminosity of the protostar remains low. Hence at earlier times these effects may be reduced. Secondly, while the densities and temperatures in pristine molecular clouds can easily explain the formation of low-mass star-forming cores, without the input of some stabilising energy such as an external pressure, micro-turbulence \citep{mckee03}, magnetic fields \citep[e.g.][]{hennebelle11,commercon11} or radiative heating and outflows \citep[e.g.][]{krumholz12}, these conditions are not conducive to creating a ``monolithic" core that does not fragment, from which the forming massive star can accrete all of its mass. Instead, massive stars may start off embedded in smaller cores that source most of their mass from the surrounding cluster-forming clump \citep{bonnell11,myers11}, implying that massive stars can only form in clusters, not in isolation. 

Placing these theoretical concerns aside initially, and working under the hypothesis that massive stars form as a scaled-up version of low-mass star formation, in this paper we aim to probe the circumstellar environment of the massive star-forming region AFGL\,2591 by combining multi-wavelength observations and modelling to determine whether any of the signatures of its formation differ significantly from those of low-mass protostars. 

AFGL\,2591 is a well-studied example of a luminous star-forming region \citep[2.1-2.5$\times10^5\,$L$_{\odot}$ at 3.33\,kpc,][]{lada84,henning90,rygl12}. One of its most prominent features is a one sided conical reflection nebula observed in the near-IR \citep[e.g][]{minchin91,tamura92}. Projected within the Cygnus-X star-forming complex, the distance to the source has recently been determined by trigonometric parallax measurements to be 3.33$\pm$0.11\,kpc, more distant than previously assumed \citep[between 1 and 2\,kpc, e.g.][]{poetzel92,hasegawa95,trinidad03,van-der-tak05}. As will be described below, AFGL\,2591 actually consists of several objects. However, as one source, AFGL\,2591-VLA\,3, dominates the SED and infrared images and hence the luminosity, the name AFGL\,2591 will therefore also be used henceforth to refer to this dominant source.

AFGL\,2591 has been studied and modelled by many authors. For example, one-dimensional modelling of the circumstellar geometry via the observed SED has been carried out by \citet{guertler91,van-der-tak99,mueller020} and \citet{de-wit09}. Improving on these, \citet{preibisch03} used two models - one of a disk and the other of an envelope with outflow cavities - to reproduce the 40\,AU diameter (assuming d=1\,kpc) bright disk of emission observed in their K band image, and \citet{trinidad03} have modelled the millimetre emission from AFGL\,2591-VLA\,3 as an optically thick disk without an envelope. As part of their comprehensive study, \citet{van-der-tak99} modelled their observed molecular lines (CS, HCN, HCO$^{+}$) including an outflow cavity in a power-law envelope, finding that a half-opening angle of 30$^{\circ}$ was able to better reproduce the line profiles. \citet{van-der-wiel11} have modelled a set of six lines detected toward AFGL\,2591 as part of the JCMT Spectral Legacy Survey, finding evidence of a cavity or inhomogeneity in the envelope on scales of $\le10^4$AU. In addition, as well as observing a bi-conical outflow structure in $^{12}$CO (2-1) on scales of 1-2$''$ or several thousand AU, \citet{jimenez-serra12} also uncovered evidence for chemical segregation within the inner 3000\,AU of the AFGL\,2591-VLA\,3 envelope (assuming d=3\,kpc, similar to our assumed distance of 3.33\,kpc).

The ionized gas emission in the region surrounding AFGL\,2591 has previously been observed by \citet{campbell84,tofani95,trinidad03} and \citet{van-der-tak05} from 5 to 43\,GHz. These observations showed that AFGL\,2591 is in fact not an isolated forming star, uncovering four continuum sources in the region. The observed fluxes of two of these, VLA\,1 and VLA\,2, gave spectral indices consistent with optically thin free-free emission from H{\sc II} regions, and a third source VLA\,3, was measured to have a steeper spectral index, possibly indicating optically thick emission. VLA\,3 is also coincident with the central illuminating source of AFGL\,2591 observed at shorter wavelengths \citep{trinidad03}. A fourth radio continuum source has also been detected by \citet{campbell840} and \citet{tofani95} (their source 4 and n4 respectively), which shall be referred to as VLA\,4 in the following sections. In addition, the 3.6\,cm images presented in this work uncover a fifth source, VLA\,5, to the south west.

\begin{table*}
\tiny
\centering
\caption{Summary of VLA observations}  
\begin{tabular}{@{}lllllllll@{}}
\hline
Wavelength & Configuration & Date of & Program & No. of & Time & \multicolumn{2}{l}{Pointing centre} & Phase calibrator \\
&& observation & & antennas & on-source (hr) & R.A. & Decl. & flux density (Jy) \\
\hline
3.6\,cm & A & 2007 Jul 26 & AJ337 & 26 (22) & 1.75 & 20 29 24.90 & +40 11 21.00 (J2000) & 1.49  \\
3.6\,cm & B & 2008 Jan 18 & AJ337 & 26 (21) & 1.96 & 20 29 24.90 & +40 11 21.00 (J2000) & 1.91  \\
3.6\,cm & C & 2008 Mar 9 & AJ337 & 27 (23) & 1.97  & 20 29 24.90 & +40 11 21.00 (J2000) & 2.13 \\
3.6\,cm & D & 2007 Apr 12 & AJ332 & 26 (20) & 0.37  & 20 29 24.90 & +40 11 21.00 (J2000) & 1.25 \\
7\,mm & A & 2002 Mar 25 & AT273 & 27 (23) & 1.97 & 20 27 35.95 & +40 01 14.90 (B1950)  & 3.44 \\
7\,mm & B & 2008 Jan 18 & AJ337 & 26 (24) & 3.43 & 20 29 24.90 & +40 11 21.00 (J2000) & 3.15 \\
7\,mm & C & 2008 Mar 9 & AJ337 & 27 (25) & 4.18 & 20 29 24.90 & +40 11 21.00 (J2000) & 4.46 \\
7\,mm & D & 2007 Apr 28 & AJ332 & 26 (17) & 0.44 & 20 29 24.90 & +40 11 21.00 (J2000) & 1.64 \\
\hline
\end{tabular}
\label{vlaobstable}
\end{table*}

Knots of H$_{2}$ and [SII] Herbig Haro objects have been detected toward AFGL\,2591 \citep{tamura92,poetzel92}, suggesting the presence of shocked gas. These are coincident with an east-west bipolar outflow \citep[e.g.][]{lada84,hasegawa95}, which extends across 5$'$ or 4.8\,pc at 3.33\,kpc, but also contains a more collimated central small-scale component with an extent of 90$'' \times$ 20$''$.

Evidence for the presence of a disk or rotationally flattened material around the central source of AFGL\,2591 has been found by several authors. The near-IR imaging polarimetry observations of \citet{minchin91} showed that a disk or toroid of material was needed to appropriately scatter the emission. In addition, a large (50$'' \times$ 80$''$) flattened ``disk'' of material has been observed perpendicular to the outflow in observations of CS lines \citep{yamashita87}. At smaller scales, \citet{van-der-tak06} and \citet{wang12} found evidence for a disk of diameter 800\,AU at 1\,kpc (corresponding to 2700\,AU at 3.33\,kpc), which exhibits a systematic velocity gradient in the northeast-southwest direction. This gradient was also found in the SMA observations of \citet{jimenez-serra12}, which they found to be consistent with Keplerian-like rotation around a 40\,M$_{\odot}$ star. Finally, both OH and water masers have been observed towards VLA\,3 \citep{trinidad03,hutawarakorn05,sanna12}. The Very Large Array (VLA)\footnote{Now the Jansky Very Large Array} 22\,GHz water maser observations of \citet{trinidad03} uncovered a maser cluster, which included a $\sim$0.01$''$ diameter shell-like structure on the smallest scales. As well as finding largely consistent results, the Very Long Baseline Array 22\,GHz water maser observations of \citet{sanna12} determined that the maser cluster is arranged in a v-shape, which coincides with the expected location of the outflow walls apparent in the near-IR reflection nebula.

In this paper, we have adopted a multi-wavelength approach to further probe the circumstellar environment of AFGL\,2591, and to address the question of whether it forms in a similar manner to its low-mass counterparts. Building on previous work, the modelling presented in this paper includes the first simultaneous radiative transfer model of the near-IR through sub-mm SED, and near-IR images, of AFGL\,2591 with a three-dimensional axisymmetric geometry. In addition, we present new multi-configuration VLA 3.6\,cm and 7\,mm continuum observations that for the first time trace an ionized jet at 3.6\,cm, as well as $^{13}$CO(1-0), C$^{18}$O(1-0) and 3\,mm continuum Combined Array for Research in Millimeter-wave Astronomy (CARMA\footnote{Support for CARMA construction was derived from the states of California, Illinois, and Maryland, the James S. McDonnell Foundation, the Gordon and Betty Moore Foundation, the Kenneth T. and Eileen L. Norris Foundation, the University of Chicago, the Associates of the California Institute of Technology, and the National Science Foundation. Ongoing CARMA development and operations are supported by the National Science Foundation under a cooperative agreement, and by the CARMA partner universities.}) observations, which trace previously unstudied scales within the molecular outflow and envelope, to derive a self-consistent picture of AFGL 2591.

Section \ref{obs_afgl} outlines the new observations carried out in this work, describing both the VLA 3.6\,cm and 7\,mm, as well as the CARMA $^{13}$CO, C$^{18}$O and 3\,mm continuum observations. Section \ref{archive} presents the archival data used in this paper, including near-IR to sub-mm SED, 2MASS photometry, measured 2MASS brightness profiles and Gemini North near-IR images of AFGL\,2591. Section \ref{sedmodelling} describes the modelling of the near-IR to sub-mm SED and near-IR images. Section \ref{result} presents the results for the SED and near-IR image modelling, as well as for the centimetre and millimetre wavelength datasets. Section \ref{discussion} presents our discussion, which covers the topics of the jet and outflow of VLA\,3, and the properties of AFGL\,2591 as a cluster. Our conclusions are given in Section~\ref{afgl:conclusions}.

\section{Radio Interferometric Observations}
\label{obs_afgl}

\subsection{VLA 3.6\,cm and 7\,mm Continuum}
\label{obscm}
Multi-configuration radio continuum observations at 3.6\,cm and 7\,mm were conducted between April 2007 and March 2008 with the VLA of the National Radio Astronomy Observatory\footnote{The National Radio Astronomy Observatory is a facility of the National Science Foundation operated under cooperative agreement by Associated Universities, Inc.}. During this time frame, the VLA continuum mode consisted of four 50\,MHz bands, two of which were placed at 8.435\,GHz, and two at 8.485 GHz for 3.6\,cm, and similarly two at 43.315\,GHz and two at 43.365 GHz for 7\,mm. At 3.6\,cm, observations were taken in all four configurations of the VLA (A to D), and at 7\,mm, observations were performed in B through D array and combined with A array archive data previously published in \citet{van-der-tak05}. When combined, these observations provided baseline lengths between 35\,m and 36.4\,km, giving information on angular scales from 0.24$''$ to 3$'$ for 3.6\,cm, and 0.05 to 43$''$ for 7\,mm. 

For each observation, Table \ref{vlaobstable} lists the observed wavelength, configuration, observation date, program code, number of antennas in the array (with the number of antennas with useful data shown in brackets), time on-source, the pointing centre of the target, and the flux density of the gain calibrator determined from bootstrapping the flux from the primary flux calibrator. The number of antennas with useful data was reduced due to antennas being out of the array for upgrading and testing of new VLA capabilities, general hardware issues, and also some residual bad data. For all observations, the gain calibrator was 2015+371 and primary flux calibrator was 1331+305 (3C286). 

Data reduction and imaging were carried out using the Common Astronomy Software Applications (CASA)\footnote{http://casa.nrao.edu} package. As 1331+305 was slightly resolved, a model was used for flux calibration, which has a flux of 5.23\,Jy at 3.6 cm and 1.45\,Jy at 7\,mm, allowing data at all \emph{uv} distances to be used. Antenna gain curves and an opacity correction (zenith opacity = 0.06) were applied for the 7\,mm data. The error in absolute flux calibration is approximately 1-2\% for the 3.6\,cm band, and 3-5\% for the 7\,mm band. The data were imaged using a CLEAN multi-scale deconvolution routine, using three scales, and Briggs weighting with a robust parameter of 0.5. The rms noise in the final combined images were 30 and 56\,$\mu$Jy\,beam$^{-1}$ for the 3.6\,cm and 7\,mm images respectively, and the synthesised beams were 0.43$''$ by 0.40$''$, P.A.=43$^{\circ}$ and 0.11$''$ by 0.11$''$, P.A.=43$^{\circ}$. These values, as well as the synthesised beam and map rms for each part of the CARMA observations described in Section \ref{obsmm}, are summarised in Table~\ref{allobstable}.

\subsection{CARMA $^{13}$CO, C$^{18}$O and 3mm Continuum}
\label{obsmm} 
CARMA observations at $\lambda\sim3$\,mm were taken on 10 July 2007 in D configuration (corresponding to baseline lengths between 11-150\,m). A total of 14 antennas were used during the observations, five 10-m and nine 6-m in diameter. 

The CARMA correlator was comprised of two side bands, placed either side of the chosen LO frequency, which for these observations was 108.607\,GHz. Both the upper and lower side bands contained three spectral windows: one wide and two narrow, giving a total of six windows. The two wide spectral windows, centred on 106.7 and 110.5\,GHz, had a bandwidth of 500\,MHz and a total of 15 channels, and the four narrow spectral windows had a width of approximately 8\,MHz and 63 channels, giving a spectral resolution of 122\,kHz. The observed lines, $^{13}$CO(J=1-0) and C$^{18}$O(J=1-0), lay in the two narrow spectral windows in the upper side band at 110.201 and 109.782\,GHz. At these frequencies, the spectral resolution was $\sim$0.33\,km\,s$^{-1}$. 

The phase centre for AFGL\,2591 was 20$^{\rm h}$29$^{\rm m}$24$^{\rm s}$.7 +40$^{\circ}$11$'$18$''$.9 (J2000), and the total time on source was 2.5\,hr. MWC\,349 was used as a gain and bandpass calibrator, and Neptune as a flux calibrator. Data calibration and imaging were carried out in CASA. To do this, the data was exported to CASA from its original MIRIAD format after applying line length corrections and Hanning smoothing. Using a flux for Neptune of $\sim$4.2\,Jy at 108\,GHz (W. Kwon, private communication), fluxes of 1.2 and 1.3\,Jy for MWC\,349 were derived for the upper and lower side bands respectively. As there was insufficient signal-to-noise on MWC\,349 in the narrow spectral windows, band pass calibration was only performed on the two wide spectral windows. However, as the amplitude bandpass shape varies across the 8\,MHz band by less than 3\%, and similarly for the phase (D. Friedel, private communication), this should not significantly affect our results. To image the data, Briggs weighting and a robust parameter of 0.5 were used. The beam size and rms noise for each observed line and continuum spectral window are given in Table \ref{allobstable}.

\begin{table}[!htbp]
\tiny
\centering
\caption{Summary of radio interferometric observations}  
\begin{tabular}{@{}llllll@{}}
\hline
$\lambda$ & Array & line/ & \multicolumn{2}{l}{Synthesised beam} & Map rms \\
& & cont. & Size ($''$) & P.A. ($^{\circ}$) & (mJy\,beam$^{-1}$)  \\
\hline
3.6\,cm & VLA-A to D & cont. &  0.43$\times$0.40 & 43  & 0.030 \\
7\,mm & VLA-A to D & cont. & 0.11$\times$0.11 & 43 & 0.056 \\
3\,mm & CARMA-D &  $^{13}$CO & 4.4$\times$3.7 & 96 & 100 \\
3\,mm & CARMA-D &  C$^{18}$O & 4.5$\times$3.6 & 93 & 100 \\
2.8\,mm & CARMA-D & cont. & 4.9$\times$4.1& -4.8 & 2.2 \\
2.7\,mm & CARMA-D & cont. & 4.3$\times$3.5 & 95 & 2.1 \\
\hline
\end{tabular}
\label{allobstable}
\end{table}

\section{Archival Data}
\label{archive}
\subsection{SED}
\label{SEDnearIRafgl}
\begin{figure}[!htbp]
\sidecaption
\includegraphics[width=9cm]{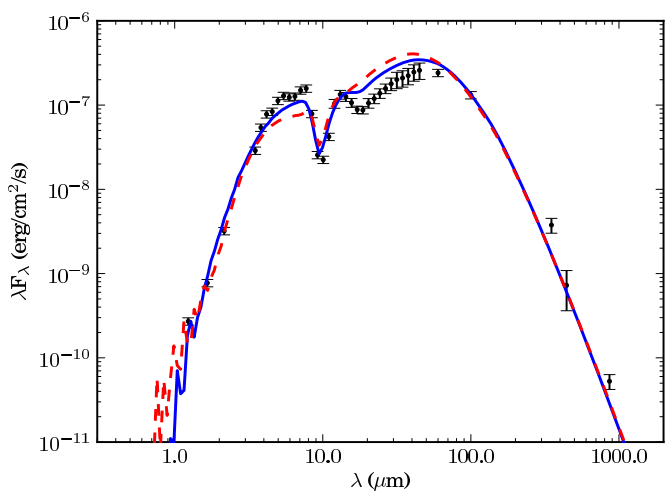}
 \caption[]{The SED of AFGL\,2591, collated from the literature. The best-fitting models for an envelope with and without a disk (overplotted solid blue and dashed red lines respectively) are discussed in Section~\ref{sedresults_afgl}.
The errors shown are those reset to 10\% if the error in the measured flux was~$<$10\%.}
 \label{seddata}
\end{figure}

The observed near-IR to sub-millimetre SED of AFGL\,2591 is shown in Fig. \ref{seddata}. The SED data points were collected from the literature; Table \ref{obstable} lists the wavelengths, flux densities and references for the data displayed in Fig.~\ref{seddata}. The 30 data points shown between 3.5 and 45\,$\mu$m were sampled uniformly in log-space from a highly processed ISO-SWS spectra of AFGL\,2591 \citep{sloan03} taken on 7 Nov 1996, observation ID 35700734. The uncertainties for the averaged ISO-SWS spectrum were assumed to range from 4 to 22\% as per the ISO handbook volume~V\footnote{http://iso.esac.esa.int/users/handbook/}. As mentioned in Section \ref{intro}, the SED and therefore flux densities of AFGL\,2591 given in Table \ref{obstable} were assumed to be dominated by one source. Evidence to support this assumption will be presented in Section \ref{vla1}, via a comparison of the SEDs of the resolved sources in the region.

\begin{table*}
\center
  \caption[]{Observed near-IR to sub-millimetre fluxes for AFGL\,2591, collated from the literature.}  
  \begin{tabular}{@{}llllll@{}}
  \hline
  Wavelength & Flux Density & Aperture & Description / Origin \\
 ($\mu$m) & (Jy) & radius & \\
 \hline
1.235 & 0.11 $\pm$ 10\% & Irregular, $<$ 40$''$ & 2MASS J band (1) \tablefootmark{a}\\
1.662 & 0.43 $\pm$ 10\% & Irregular, $<$ 40$''$ & 2MASS H band (1) \tablefootmark{a} \\
2.159 & 2.30 $\pm$ 10\% & Irregular, $<$ 40$''$ & 2MASS K$_{\rm s}$ band (1) \tablefootmark{a} \\
 3.500 &  33.68 $\pm$ 1.347 & ISO apertures & Sampled ISO-SWS spectrum (2) \\
 3.822 &  68.98 $\pm$ 2.759 & ISO apertures & Sampled ISO-SWS spectrum (2) \\
 4.174 &  108.4 $\pm$ 7.590 & ISO apertures & Sampled ISO-SWS spectrum (2) \\
 4.558 &  127.2 $\pm$ 8.906 & ISO apertures & Sampled ISO-SWS spectrum (2) \\
 4.978 &  186.8 $\pm$ 13.08 & ISO apertures & Sampled ISO-SWS spectrum (2) \\
 5.436 &  233.8 $\pm$ 16.37 & ISO apertures & Sampled ISO-SWS spectrum (2) \\
 5.937 &  245.0 $\pm$ 17.15 & ISO apertures & Sampled ISO-SWS spectrum (2) \\
 6.483 &  273.7 $\pm$ 19.16 & ISO apertures & Sampled ISO-SWS spectrum (2) \\
 7.080 &  353.2 $\pm$ 24.73 & ISO apertures & Sampled ISO-SWS spectrum (2) \\
 7.732 &  406.2 $\pm$ 28.43 & ISO apertures & Sampled ISO-SWS spectrum (2) \\
 8.444 &  221.6 $\pm$ 15.51 &ISO apertures & Sampled ISO-SWS spectrum (2) \\
 9.221 &  78.69 $\pm$ 5.508 &ISO apertures & Sampled ISO-SWS spectrum (2) \\
10.070 & 75.49 $\pm$ 5.285 &ISO apertures & Sampled ISO-SWS spectrum (2) \\
10.997 & 154.3 $\pm$ 10.80 & ISO apertures & Sampled ISO-SWS spectrum (2) \\
12.009 & 416.8 $\pm$ 50.01 &ISO apertures & Sampled ISO-SWS spectrum (2) \\
13.115 & 583.4 $\pm$ 70.01 &ISO apertures & Sampled ISO-SWS spectrum (2) \\
14.322 & 600.0 $\pm$ 72.00 &ISO apertures & Sampled ISO-SWS spectrum (2) \\
15.641 & 556.8 $\pm$ 66.81 &ISO apertures & Sampled ISO-SWS spectrum (2) \\
17.081 & 503.2 $\pm$ 50.32 & ISO apertures & Sampled ISO-SWS spectrum (2) \\
18.653 & 542.3 $\pm$ 54.23 &ISO apertures & Sampled ISO-SWS spectrum (2) \\
20.370 & 717.7 $\pm$ 93.31 &ISO apertures & Sampled ISO-SWS spectrum (2) \\
22.245 & 886.1 $\pm$ 115.2 &ISO apertures & Sampled ISO-SWS spectrum (2) \\
24.293 & 1118 $\pm$ 145.3 &ISO apertures & Sampled ISO-SWS spectrum (2) \\
26.530 & 1391 $\pm$ 180.9 &ISO apertures & Sampled ISO-SWS spectrum (2) \\
28.972 & 1728 $\pm$ 293.8 &ISO apertures & Sampled ISO-SWS spectrum (2) \\
31.639 & 2118 $\pm$ 465.9 &ISO apertures & Sampled ISO-SWS spectrum (2) \\
34.552 & 2413 $\pm$ 530.9 &ISO apertures & Sampled ISO-SWS spectrum (2) \\
37.733 & 2810 $\pm$ 618.3 &ISO apertures & Sampled ISO-SWS spectrum (2) \\
41.207 & 3384 $\pm$ 744.5 &ISO apertures & Sampled ISO-SWS spectrum (2) \\
45.000 & 3879 $\pm$ 853.5 &ISO apertures & Sampled ISO-SWS spectrum (2) \\
60 & 4830 $\pm$ 290 & 6$'$ & IRAS 60$\mu$m (3) \tablefootmark{b} \\
100 & 4390 $\pm$ 290 & 6$'$ & IRAS 100$\mu$m (3) \tablefootmark{b} \\
350 & 440 $\pm$ 20\% & 60$''$ & CSO (4) \\
450 & 107 $\pm$ 50\% & 60$''$ & JCMT (5) \tablefootmark{c} \\
850 & 15 $\pm$ 20\% & 72$''$ & JCMT (5) \tablefootmark{c} \\
\hline
\end{tabular}
\tablefoot{
\tablefoottext{a}{Image shown in Fig. \ref{JHKrgb}. Fluxes found by irregular aperture photometry.}
\tablefoottext{b}{Fluxes found using radial aperture photometry on IRIS coadded images.}
\tablefoottext{c}{Fluxes found using radial aperture photometry on SCUBA Legacy Catalogue images.}
}
\tablebib{(given in parentheses in the column \textit{Description / Origin): (1) \citet{skrutskie06}; (2) \citet{sloan03}; (3) \citet{miville-deschenes05}; (4) \citet{mueller02}; (5) \citet{di-francesco08}. }}
\label{obstable} 
\end{table*}

\subsection{Near-IR 2MASS Images}
\label{pres2MASS}
\begin{figure*}[!htbp]
\includegraphics[width=18cm]{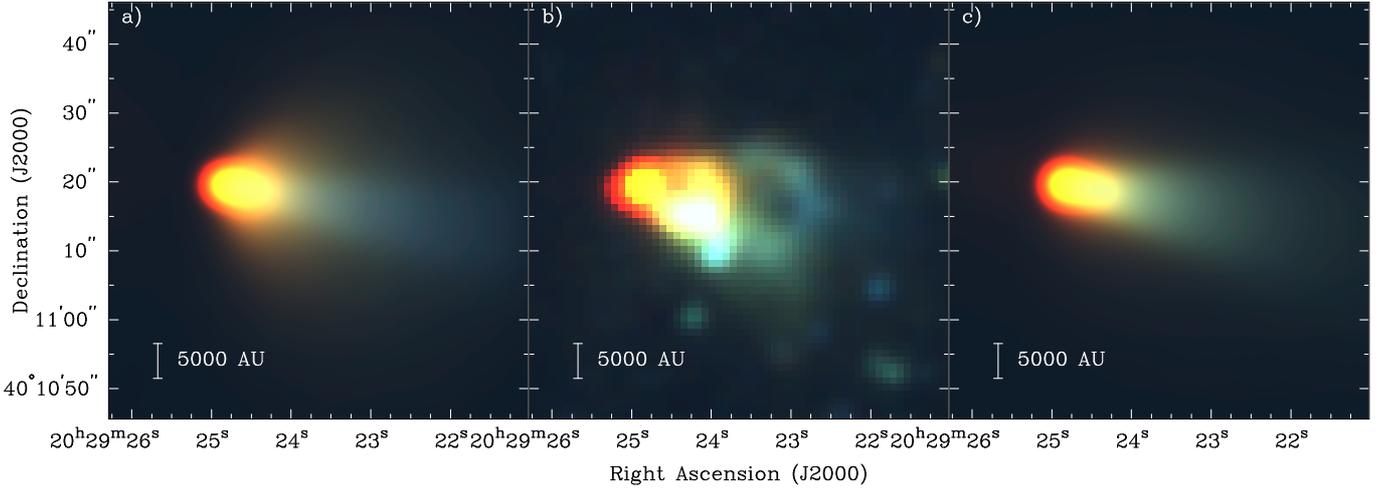}
 \caption[Observed and model J, H and K$_{\rm s}$ band three-colour images]{
a) Model three-colour J, H and K$_{\rm s}$ band image for the envelope with disk model (RGB: K, H, and J bands respectively). b) Observed 2MASS three-colour image of AFGL\,2591. c) Model J, H, and K$_{\rm s}$ band image for the envelope without disk model. The model images have been normalised to the total integrated fluxes given in Table \ref{obstable}, in order that the morphology of the emission can be easily compared. Stretch: red: K$_{\rm s}$ band, 130-300 MJy\,sr$^{-1}$; green: H band, 95-150 MJy\,sr$^{-1}$, blue: J band, 30-60 MJy\,sr$^{-1}$.}
 \label{JHKrgb}
\end{figure*}

A three-colour 2MASS \citep{skrutskie06} J, H, and K$_s$ band image of the near-IR emission observed towards AFGL\,2591 is shown in the middle panel of Fig. \ref{JHKrgb}, where red, green and blue are K$_s$, H and J bands respectively. This Figure shows the conical near-IR reflection nebula  \citep[previously observed by e.g.][]{kleinmann75,minchin91,tamura92}, which is consistent with being a blue-shifted outflow lobe from the central source of AFGL\,2591.

To obtain integrated J, H and K$_s$ band fluxes for AFGL\,2591, irregular aperture photometry was performed on 2MASS Atlas images. Photometric uncertainties on the fluxes are $<$1\%, however it is likely the fluxes are more uncertain due to the choice of the irregular photometry apertures, and foreground or background stars being included within them. The uncertainty due to contaminating stars is likely dominated by the 2MASS source 20292393+4011105 at 20$^{\rm h}$29$^{\rm m}$23$^{\rm s}$.93 +40$^{\circ}$11$'$10$''$.56 (J2000; easily visible as a point source in the Gemini North near-IR images presented in Section \ref{obsresultcm}), as the apertures were placed to avoid all other point sources visible in the 2MASS images. The fluxes of this contaminating source are 0.011, 0.035 and 0.105\,Jy at J, H and K$_s$ bands respectively, with K$_s$ band being an upper limit. Therefore, the uncertainty in the measured 2MASS fluxes of the AFGL\,2591 outflow lobe was estimated to be $\sim$10\%.

\subsection{2MASS Brightness Profiles}
Flux profiles of AFGL\,2591, summed across strips aligned with and perpendicular to the source outflow axis (P.A.=256$^{\circ}$, taken from \citet{sanna12}, with thicknesses of 24.6$''$ and 104$''$ respectively), were measured for the three 2MASS bands. The background subtracted, normalised profiles are shown in Fig.~\ref{fig:afglprofiles} along with the best-fitting models to both the SED and profiles, for the models consisting of an envelope with and without a disk (blue solid and red dashed lines respectively), which will be discussed further in Section \ref{sedresults_afgl}. The background was determined by finding the average value within two strips either side of the main profile. The errors in the profiles shown in each panel of Fig.~\ref{fig:afglprofiles} reflect the uncertainty due to background fluctuations, and are calculated as the standard deviation of the profiles measured in the two background strips, which were assumed to contain minimal source flux. As the strips used to calculate the uncertainty in the perpendicular profiles contained several bright stars, iterative sigma clipping was performed, i.e. pixels lying more than 5 standard deviations away from the mean were then ignored and the mean and standard deviation were recalculated and the process was repeated until no pixels were rejected.

\subsection{Near-IR Gemini North NIRI Images}
\label{gemini}
High resolution Near-IR images of AFGL\,2591 were obtained from the Gemini website, which were taken under program GN-2001A-SV-20 as part of the commissioning of Gemini North's NIRI instrument, and were available under public release\footnote{http://www4.cadc-ccda.hia-iha.nrc-cnrc.gc.ca/gsa/sv/dataSVNIRI.html}. The total integration time was 2 min for J band and 1 min at H and K$'$ bands. We aligned the images using Chandra X-ray sources in the field \citep{evans10}, giving a positional accuracy of 0.6$''$, and calibrated the flux level of the images in MJy\,sr$^{-1}$ using aperture photometry of bright sources in both the Gemini North and 2MASS images. The FWHMs of the PSFs in the images range between 0.3-0.4$''$. The peak position of the central source of AFGL\,2591 in the J band image was found to be 20$^{\rm h}$29$^{\rm m}$24$^{\rm s}$.86 +40$^{\circ}$11$'$19$''$.5 (J2000), which was taken to be the position of the powering object. As these images were saturated in K$'$ band, 2MASS images were instead used for the modelling described in Section \ref{sedmodelling}. However, in Sections \ref{obsresultcm} and \ref{obsresultmm} these high-resolution images are compared to the radio continuum and molecular line data observed for this work.

\section{SED and Near-IR Image Modelling}
\label{sedmodelling}

\subsection{The Dust Continuum Radiative Transfer Code}
\label{dustcode_afgl}

The SED of AFGL\,2591 was modelled between near-IR and sub-mm wavelengths using the 3D Monte Carlo dust radiative transfer code \textsc{Hyperion} described in \citet{robitaille11} and the genetic search algorithm used to model IRAS\,20126+4104 in \citet{johnston11}. 

For a given source and surrounding dust geometry sampled in a spherical-polar grid, \textsc{Hyperion} models the nonisotropic scattering and thermal emission by/from dust, calculating radiative equilibrium dust temperatures \citep[using the technique of][]{lucy99}, and producing spectra and multi-wavelength images. As dust and gas are assumed to be coupled in the code we are able to probe the bulk of the material which surrounds AFGL\,2591. In this section, we describe the density structure of the disk, envelope and outflow cavity in the model.

We model the circumstellar geometry of AFGL\,2591 with the same 3D axisymmetric envelope with disk geometry successfully employed to model the SEDs and scattered light images of low-mass protostars \citep[e.g.][]{robitaille07,wood02a,whitney03}. The axisymmetric three dimensional flared disk density is described between inner and outer radii $R^{\rm min}$ and $R_{\rm disk}$ by 
\begin{equation}
\rho_{\rm disk} (\varpi,z)=\rho_0 \left (\frac{R_0}{\varpi}\right )^{\alpha}
\exp{ \left\{ -{1\over 2} \left[\frac{z}{h( \varpi )}\right]^2 \right\} }
\; ,
\label{discdensity}
\end{equation}
where $R_{0}$=100\,AU, $\varpi$ is the cylindrical radius, $z$ is the height above the disk midplane and $\rho_0$ is set by the total disk mass $M_{\rm disk}$, by integrating the disk density over $z$, $\varpi$ the cylindrical radius, and $\phi$ the azimuthal angle. The scaleheight $h(\varpi)$ increases with radius: $h(\varpi)=h_0\left ( {\varpi /{R_0}} \right )^\beta$ where $h_0$ is the scaleheight at $R_0$. We have assumed the parameter $\beta=1.25$, derived for irradiated disks around low mass stars \citep{dalessio98}, and have taken $\alpha=2.25$, to provide a surface density of $\Sigma\sim \varpi^{-1}$. This equation is essentially the same as equation (1) given in \citet{johnston11}, however their second term is removed as this pertains to accreting $\alpha$-disks, whereas our disk is not accreting. Further, equation (\ref{discdensity}) is expressed in terms of $R_0$ instead of $R_\star$ so that the scaleheight is defined at a radius which is more conceptually accessible.

The density of the circumstellar envelope is taken to be that of a rotationally flattened collapsing spherical cloud 
\citep[][]{ulrich76,terebey84} with radius $R^{\rm max}_{\rm env}$,
\begin{equation}
\rho_{\rm env} (r,\mu) = \rho_{{\rm env}0} 
\left( {{r}\over{R_c}}\right)^{-3/2}\left(1+{{\mu}\over{\mu_0}} \right)^{-1/2} \left( {{\mu}\over{\mu_0}}+{{2\mu_0^2R_c}\over{r}}\right)^{-1}
\end{equation}
where $\rho_{{\rm env}0}$ is the density scaling of the envelope, $r$ is the spherical radius, $R_c$ is the centrifugal radius, $\mu$ is the cosine of the polar angle ($\mu=\cos\theta$), and $\mu_0$ is the cosine of the polar angle of a streamline of infalling particles in the envelope as $r\rightarrow\infty$. The equation for the streamline is given by
\begin{equation}
\mu_0^3 + \mu_0(r/R_c-1)-\mu(r/R_c)=0\; 
\end{equation}
which can be solved for $\mu_0$. In our model, we assume the centrifugal radius is also the radius at which the disk forms $R_c~=~R_{\rm disk}$. 

We chose to describe the envelope density via a density scaling factor instead of the envelope accretion rate and stellar mass used in \citet{johnston11}, as this does not require the assumption of evolutionary tracks - which are currently very uncertain for massive stars - to provide a physical stellar radius and temperature for a given stellar mass. Instead, we have varied the stellar radius and temperature as parameters (described in Section \ref{input}). 

To reproduce the morphology of the emission in the near-IR images, we also include a bipolar cavity in the model geometry with density $\rho_{\rm cav}$, and a shape described by
\begin{equation}
z(\varpi) = z_0 \varpi^{b_{\rm cav}}
\label{cavity}
\end{equation}
where the shape of the cavity is determined by the parameter $b_{\rm cav}$ which we set to be $b_{\rm cav}=1.5$, and $z_0$ is chosen so that the cavity half-opening angle at $z=10,000$\,AU is $\theta_{\rm cav}$. The cavity was included by resetting the envelope density to $\rho_{\rm cav}$ inside the region defined by equation (\ref{cavity}) only where the envelope density was initially larger. 

The inner radius of the dust disk and envelope $R^{\rm min}$ is expressed in terms of the dust destruction radius $R_{\rm sub}$, which was found empirically by \citet{whitney04c} to be 
\begin{equation}
R_{\rm sub} = R_{\star}(T_{\rm sub}/T_{\star})^{-2.1}
\end{equation}
where the temperature at which dust sublimates is assumed to be  $T_{\rm sub}=1600$K, and $T_{\star}$ is the temperature of the star.
 
For the genetic algorithm, which is described in detail in Section 4.3 of \citet{johnston11}, the size of the first generation $N$ was set to 1000, and the size of subsequent generations $M$ was set to be 200. The code was taken to be converged when the $\chi^2$ value of the best fitting model decreased less than 5\% in 20 generations. 

\begin{figure*}[!htbp]
\includegraphics[width=18cm]{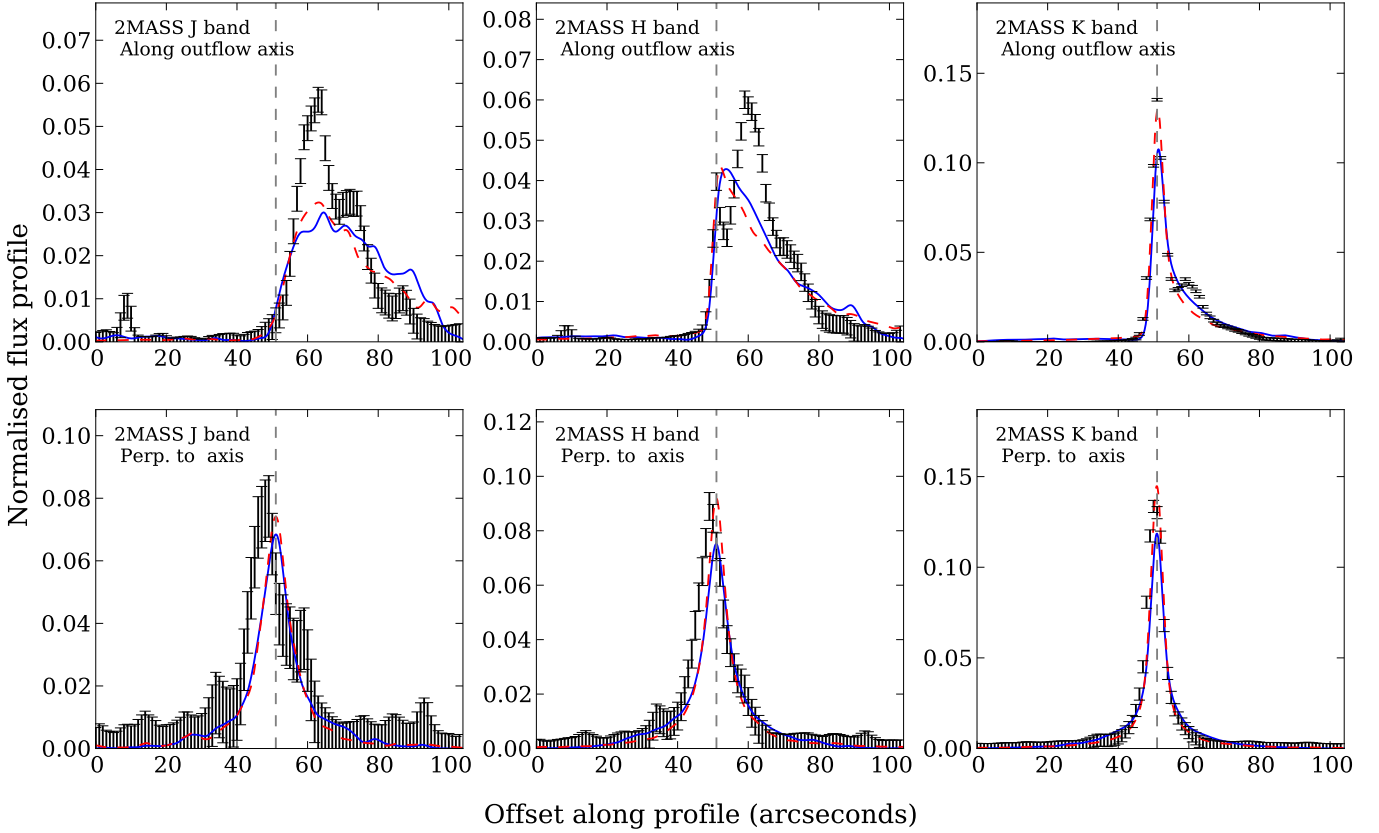}
 \caption[Normalised observed and model flux profiles for the three 2MASS bands]{Black error bars: normalised flux profiles for the three 2MASS bands, aligned with (top) and perpendicular to (bottom) the outflow axis. Blue and red lines: the profiles of the best-fitting models to both the observed SED and profiles, for envelope with disk and without disk models respectively. Vertical grey dashed lines mark the position of the central source.}
 \label{fig:afglprofiles}
\end{figure*}

\subsection{Input Assumptions: Model Parameters}
\label{input}

A set of plausible ranges for parameters describing the model, in which the genetic algorithm searched for the best fit, is given in Table \ref{ranges}. In addition to this, the stellar radius and temperature were also required to lie above the zero-age main sequence defined as $\log{T_{\rm ZAMS}}$ vs. $\log{(0.9\times R_{\rm ZAMS})}$ for the solar metallicity models of \citet{schaller92}, where the factor of 0.9 is to ensure that enough models are sampled around the ZAMS, rather than strictly above the ZAMS. The stellar temperature $T_{\star}$ and a stellar surface gravity of $\log{g} = 4$ were then used to select a model from the stellar atmosphere grid of \citet{castelli04}. To sample the cavity half-opening angle $\theta_{\rm cav}$, we first determined the relation between the half-opening angle and the inclination such that the equation of the cavity, equation (\ref{cavity}), projected onto the plane of the sky produced the right opening angle observed in the Gemini near-IR images,

\begin{equation}
i = \arcsin{\left[  \frac{ 0.826~(\sin{\theta_{\rm cav}})^b}{ \cos{\theta_{\rm cav}}}\right]}.
\end{equation}

When sampling the half-opening angle, the inclination was first sampled and equation (6) was then used to determine an upper bound for $\theta_{\rm cav}$, as radiative transfer effects only work to make the cavity look larger than it actually is. For instance, light can be scattered into the envelope making it look larger, whereas because the density of the cavity is constant, there is no definite edge of emission seen within the cavity. The largest half-opening angle possible, for an inclination of 90$^{\circ}$, was therefore 54$^{\circ}$ at a radius of 10,000\,AU.

\begin{table*}
 \centering
  \caption{Assumed ranges for model parameters as input to the genetic search algorithm.}
  \begin{tabular}{@{}llll@{}}
  \hline
 Parameter     & Description & Value/Range & Sampling \\
 \hline 
 $R_{\star}$ & Stellar radius (R$_{\odot}$) & 0.6 -- 500 & logarithmic\\
 $T_{\star}$ & Stellar temperature (K) & 5000 -- 50,000 & logarithmic\\
 $R^{\rm min}$ & Envelope and disk inner radius (R$_{\rm sub}$) & 1 -- 100 & logarithmic \\
$R^{\rm max}_{\rm env}$ & Envelope outer radius (AU) & $10^{4}$ -- $10^{6}$ & logarithmic \\
$\rho_{\rm{env}0}$ & Envelope density scaling factor (g\,cm${}^{-3}$) & $10^{-21}$ -- $10^{-15}$ & logarithmic \\
$R_{\rm disk}$ or $R_{\rm c}$ & Disk or centrifugal radius (AU) & 10 -- 10$^{5}$ & logarithmic \\
$\theta_{\rm cav}$ & Cavity half-opening angle at 10$^{4}$\,AU (degrees) & 0 -- 90 & linear \\
$i$ & Inclination w. r. t. the line of sight (degrees) & 0 -- 90 & linear \\
$\rho_{\rm cav}$& Cavity ambient density (g\,cm${}^{-3}$)  & $10^{-21}$ -- $10^{-16}$ & logarithmic \\
$M_{\rm disk}$ & Disk mass (M$_\odot$) & 0.1 -- 50 & logarithmic \\
$h_{0}$ & Disk scale-height at radius of 100\,AU (AU) & 0.2 -- 20 & logarithmic \\
  \hline
\end{tabular}
\label{ranges}
\end{table*}

\subsection{Model Fitting}
\label{fitting}

As part of the genetic algorithm used here and described in \citet{johnston11}, the models were fit to the data using the SED fitting tool of \citet{robitaille07}, where $ {A_{\rm v}}$, the external extinction, was a free parameter in the fit.
The distance was set to be 3.33\,kpc, and the visual extinction ${A_{\rm v}}$ was allowed to vary between 0 and 40 magnitudes. 

As described in \citet{robitaille07}, the fitting automatically takes into account the different circular aperture sizes when fitting the models. However, in the case of the ISO fluxes, the ISO apertures are not circular, so instead the fluxes were determined by computing images at the ISO wavelengths, and summing up the flux weighted by the slit transmission maps (found within the Observer's SWS Interactive Analysis software, OSIA\footnote{http://sws.ster.kuleuven.ac.be/osia/}) which give the transmission as a function of location on the detector. The ISO slits were rotated to the correct orientation for the observation and applied to the model. During the SED fitting, each data point was weighted by the distance in wavelength between its two adjacent data points. This was done to allow the algorithm to search for a model which reproduced the shape of the entire SED, not only the sections with a high density of data. 

The models were also simultaneously fit to the six flux profiles shown in Fig. \ref{fig:afglprofiles}, measured from the 2MASS images. The model flux profiles were found in the same way as the observed profiles, by summing the flux in the model images within 24.6$''$ and 104$''$ wide strips aligned with and perpendicular to the outflow axis, and normalising them to the total flux. To produce the convolved model J,H and K$_s$ band images, the model images were convolved with the FWHM of the 2MASS point-spread functions (PSF) for each band: 3.15$''$, 3.23$''$ and 3.42$''$ for J, H and K$_s$ bands respectively, which were determined by fitting Gaussian PSFs to stars in the observed images. The model images shown in Fig. \ref{JHKrgb} are high signal-to-noise versions of the images used for fitting. 

For both the SED and profiles, the errors were reset to 10\% before fitting if the uncertainty in the measured flux was $<$10\%. This was done to take into account non-measurement errors such as variability. The combined reduced $\chi^2$ for each model given as input to the genetic algorithm was calculated as
\begin{equation}
\chi^2_{\rm combined}= \chi_{\rm SED}^2 + \chi_{\rm profiles}^2
\end {equation} 
where $\chi_{\rm SED}^2$ is the best-fitting reduced $\chi^2$ value for the SED alone, and $\chi_{\rm profiles}^2$ is the overall reduced $\chi^2$ for the profile fits.

\section{Results}
\label{result}

\subsection{Results of SED and Image Modelling}
\label{sedresults_afgl}
In this paper, we model AFGL\,2591 as a single source; as we will show in Section \ref{vla1}, the second brightest object in the region VLA\,1 does not significantly contribute to its infrared flux, therefore this is a reasonable approximation.

The genetic code was run three times - firstly with the model parameter ranges given in Table~\ref{ranges} as input for all parameters, which will be referred to as the `envelope with disk' model, and secondly with the parameter ranges given in Table~\ref{ranges} for all parameters except the disk mass and disk scale height at 100\,AU, $M_{\rm disk}$ and $h_0$, which were set to zero and were therefore not treated as model parameters. This second run will be referred to as the `envelope without disk' model. The envelope without disk model was run to ascertain whether the SED and images could be adequately produced without a disk, i.e. a simpler model which had two fewer parameters. For the envelope without disk model, $R_{\rm disk}$ will instead be referred to as the centrifugal radius of the envelope, $R_{\rm c}$, as these two parameters are interchangeable in the models. The third run had exactly the same setup as the envelope with disk model, but was used to determine the repeatability of the experiment and any parameter degeneracies, and will be referred to as the control model.

The genetic search algorithm codes were stopped after 28, 29 and 49 generations for the envelope with disk, without disk and control models respectively, when the convergence criterion was reached. This corresponds to 6545 models or 2940 CPU hours for the envelope with disk model, 6278 models or 3774 CPU hours for the envelope without disk model, and 10536 models or 3960 CPU hours for the control model.

\begin{table*}
 \centering
 \caption{Parameters of the genetic algorithm best-fitting models}
  \begin{tabular}{@{}lllll@{}}
  \hline
Parameter     & Description & Value for envelope & Value for envelope & Value for envelope with \\
& & with disk model& without disk model & disk control model \\
 \hline 
$R_{\star}$ & Stellar radius (R$_\odot$) & 90 & 200 & 52 \\[+0.03in]
$T_{\star}$ & Stellar temperature (K) & 13,000  & 8900 & 18,000 \\[+0.03in]
$R^{\rm min}$ & Envelope and disk inner radius (R$_{\rm sub}$) & 2.9 & 3.4 & 4.3 \\[+0.03in]
$R^{\rm max}_{\rm env}$ & Envelope outer radius (AU) & 180,000 & 420,000 & 790,000 \\ [+0.03in]
$\rho_{\rm{env}0}$ & Envelope density scaling factor (g\,cm${}^{-3}$) & 2.4$\times10^{-19}$ & 1.1$\times10^{-18}$ & 5.6$\times10^{-18}$ \\ [+0.03in]
$R_{\rm disk}$ or $R_{\rm c}$ & Disk or centrifugal radius (AU) & 35,000 & 9500 & 2700 \\[+0.03in]
$\theta_{\rm cav}$ & Cavity half-opening angle at 10$^{4}$\,AU (degrees) & 16 & 20 & 49 \\[+0.03in]
$i$ & Inclination w. r. t. the line of sight (degrees) & 59 & 46 & 54 \\[+0.03in]
 $\rho_{\rm cav}$& Cavity ambient density (g\,cm${}^{-3}$) & 2.1$\times10^{-20}$ & 2.0$\times10^{-20}$  & 1.0$\times10^{-21}$ \\[+0.03in]
$M_{\rm disk}$ & Disk mass (M$_\odot$) & 14 & ... & 16 \\[+0.03in]
$h_{0}$ & Disk scale-height at radius of 100\,AU (AU) & 11  & ... & 11 \\[+0.03in]
$L_{\star}$ & Stellar luminosity (L$_{\sun}$) & 2.3$\times10^{5}$ & 2.3$\times10^{5}$ & 2.4$\times10^{5}$ \\
$\rho_{\rm{env}0}~R_{\rm c}^{3/2} $ & Envelope density parameter (g\,cm${}^{-3}$ AU$^{3/2}$) & 1.6$\times10^{-12}$ & 1.0$\times10^{-12}$ & 8.0$\times10^{-13}$ \\
\hline
\end{tabular}
\label{bestfit_afgl}
\end{table*}

The resulting SEDs and profiles of the best-fitting models for the envelope with and without disk runs are shown against the data in Figs.~\ref{seddata} and \ref{fig:afglprofiles}, and the model images are compared to the observed images in Fig. \ref{JHKrgb}. The parameters of the best-fitting models (i.e. the model in each run with the lowest overall $\chi^2$ value after convergence), are also given in Table~\ref{bestfit_afgl}.

The minimum reduced $\chi^2$ for the best-fitting envelope with disk model is 14.9, and the corresponding value for the best-fitting without disk model is 18.7, therefore the model with a disk provides a marginally better fit to the SED and profiles. Comparing the model SEDs to the data shown in Fig. \ref{seddata}, it can be seen that they are qualitatively very similar, although the model with a disk fits the SED slightly better. This difference is partially caused by their disagreement in the mid-IR regime, where the without disk model does not follow the data as closely. Figure \ref{fig:afglprofiles} also shows very similar fits to the 2MASS image profiles for each model, however the without disk model fits some of the profiles marginally better, being more centrally peaked. We note that although the reduced $\chi^2$ for the best-fitting model with a disk is lower, and thus provides evidence that an envelope with an embedded disk describes the source better than a model without a disk, the difference is not large enough to prove this conclusively. We note that neither model provides a good fit to the J and H band profiles along the outflow axis. This is likely to be caused by the fact that the emission at these wavelengths is not dominated by the central source, but instead by scattered light in the surrounding medium, which in reality is highly non-uniform, as seen in Figs. \ref{JHKrgb} and \ref{cmGemfig}. Thus to reproduce the images more accurately, future modelling will have to take such inhomogeneities into account, for example the ``loops" in the outflow cavity seen in the near-IR images \citep{preibisch03}.

\begin{figure*}
\includegraphics[width=18cm]{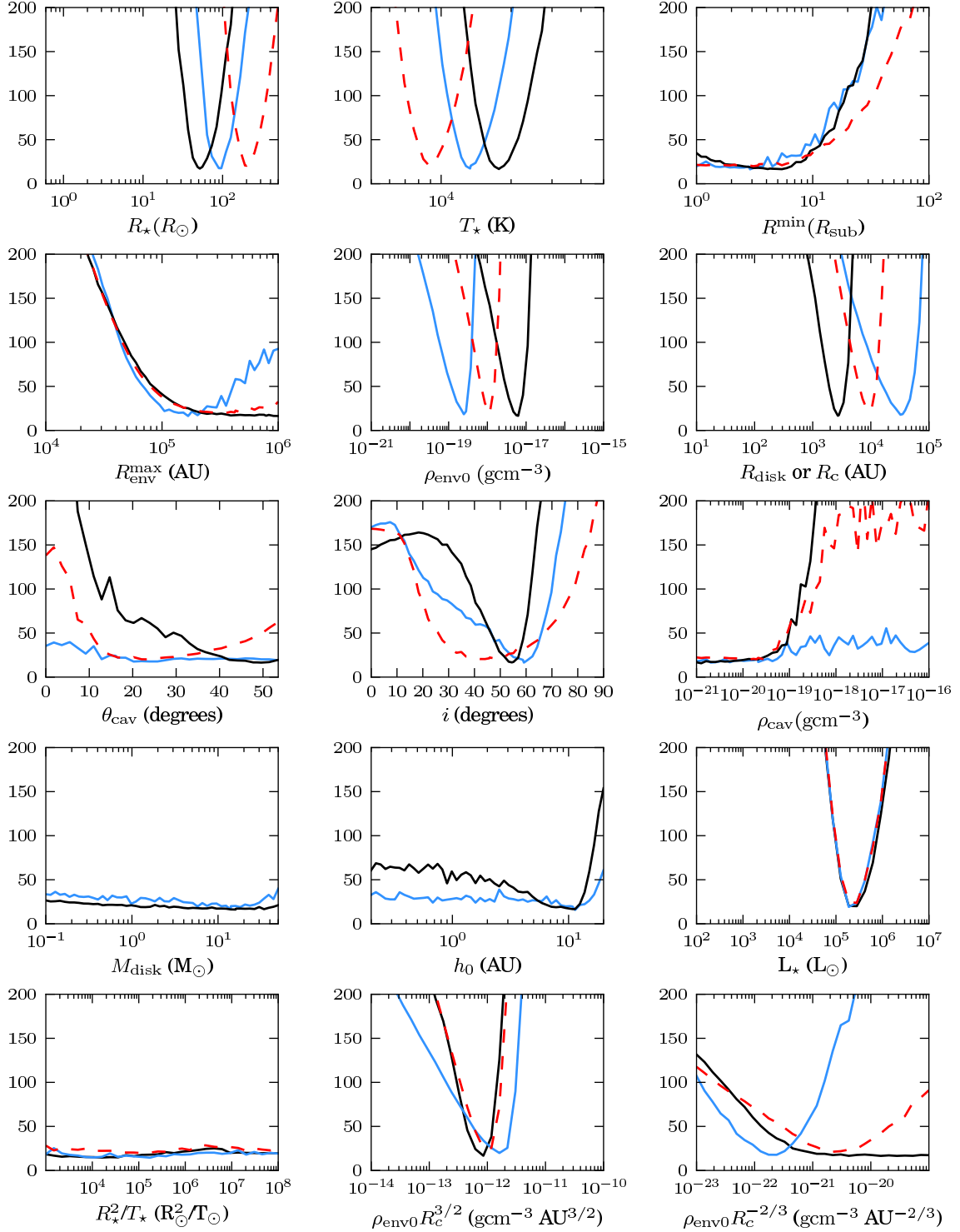}
 \caption{Plots of the $\chi^2$ surface along the axis of each model parameter for each run, determined by varying each parameter while holding the other best-fitting parameters constant. Red dashed line: envelope without disk run, light blue line: envelope with disk run, and black line: control run.}
 \label{chi2surface_afgl}
\end{figure*}

To understand how well the parameters of AFGL\,2591 were determined, we took the best-fitting models for each run and varied in turn each parameter across the original parameter ranges, while holding the other parameters constant. We then fit the resultant SED and images for each new model, and calculated their overall reduced $\chi^2$-values. From these we could therefore determine the $\chi^2$ surfaces in the vicinity of the the best-fitting model parameters and hence understand their uncertainties. Figure \ref{chi2surface_afgl} shows such a plot for each varied parameter, for all three runs. Firstly, we can see the best-fitting values and shape of the $\chi^2$ surfaces for stellar radius and temperature are different for each run, which is due to a degeneracy between these parameters. The $\chi^2$ surfaces for the stellar luminosity, which is a combination of these two parameters, is shown as a panel in Fig. \ref{chi2surface_afgl}: this surface is very well defined and very similar for all runs, showing that fitting the SED allows this parameter to be accurately and non-ambiguously determined. In the next panel we show the $\chi^2$ surface for the parameter orthogonal to the stellar luminosity, $R_{\star}^{2}T_{\star}$, is flat and thus poorly determined. Therefore the SED or image profiles do not provide a handle on either the stellar radius and temperature, only the stellar luminosity. Similarly, the envelope density scaling factor $\rho_{\rm{env}0}$ and $R_{\rm disk}$ or $R_{\rm c}$ are degenerate, as they both determine the envelope density, which can be seen from equation (2). The final two panels of Fig. \ref{chi2surface_afgl}, show the orthogonal parameter combinations $\rho_{\rm env}~R_{\rm c}^{3/2}$ and $\rho_{\rm env}~R_{\rm c}^{-2/3}$. Here $\rho_{\rm env}~R_{\rm c}^{3/2}$ is similar between the three runs, and has a sharp $\chi^2$ surface, while the best-fitting $\rho_{\rm env}~R_{\rm c}^{-2/3}$ value changes by an order of magnitude. Thus we cannot uniquely determine these two parameters from our modelling, but we can determine a parameter that combines these which governs the overall envelope density and mass, $\rho_{\rm env}~R_{\rm c}^{3/2}$.

Looking at the remaining parameters, it can be seen that although the minima in their $\chi^2$ surfaces are not sharp, the shape of their $\chi^2$ surfaces is similar for the three runs, with $R^{\rm min}$ tending to smaller values, $R^{\rm max}_{\rm env}$ tending to intermediate or large values, and the inclination preferring intermediate values. The $\chi^2$ surface for the cavity half-opening angle $\theta_{\rm cav}$ changes between each run, however is generally poorly constrained. It is also important to note that the half-opening angle is somewhat degenerate with the inclination, as the inclination of a model determines the possible upper limit for the half-opening angle as described in Section \ref{input}. The shape of the $\chi^2$ surface for the cavity density $\rho_{\rm cav}$ is variable at higher values, but always has its minimum at low values, between $10^{-21}$ and $\sim2 \times 10^{-20}$ g\,cm${}^{-3}$. For the two runs which include a disk, the $\chi^2$ surfaces of the remaining parameters $M_{\rm disk}$ and $h_{0}$ are almost flat across the sampled ranges, although they slightly prefer values close to the upper limit. However, for $h_{0}$ the $\chi^2$ rises steeply toward the upper limit of the parameter range after its minimum at 11\,AU.

In the remainder of this Section, we compare the results of our modelling to the recent results from other studies of AFGL\,2591. Firstly, our modelling finds values of roughly 2-5$R_{\rm sub}$ for the envelope and disk inner radius $R^{\rm min}$. Using equation (5), we find $R_{\rm sub}$ = 34-39\,AU for the three best-fitting models, and hence we expect values of $R^{\rm min}\sim70-200$\,AU. This value is in agreement with the radius of the inner cavity adopted by \citet{jimenez-serra12}, who in turn took this value from the disk of emission observed by \citet{preibisch03}, likely to be tracing the inner rim of the envelope and disk. At 1\,kpc, \citet{preibisch03} found the radius of this emission was 40\,AU, which is thus 130\,AU at 3.33\,kpc.

We compare the maximum envelope radius $R^{\rm max}_{\rm env}$ to the results of the JCMT line observations of \citet{van-der-wiel11}, which were able to trace the largest envelope scales, with the maximum radius in their model set to 200,000\,AU (scaled to 3.33\,kpc). This is in agreement with our SED modelling results, which prefer large envelope radii, on the scale of several hundreds of thousand AU.

We find the cavity half-opening angle $\theta_{\rm cav}$ to be poorly constrained by our modelling, although we determined an upper limit for the cavity half-opening angle of $54^{\circ}$ in Section \ref{input} (defined at 10$^4$\,AU). \citet{wang12} adopted an opening angle of 60$^{\circ}$, and hence a half-opening angle of 30$^{\circ}$, in their modelling. The models used in \citet{jimenez-serra12} and \citet{van-der-wiel11} do not include an outflow cavity.

We find the inclination to prefer intermediate values, between roughly 30-65$^{\circ}$. This is in agreement with the adopted inclination of 30$^{\circ}$ by \citet{wang12}, who in turn used the value determined by \citet{van-der-tak06}. However, \citet{jimenez-serra12} found an inclination of 20$^{\circ}$ by modelling the position-velocity diagram of combined emission from several lines. Thus, previous results appear to favour an inclination which is slightly closer to face-on. As the inclinations for the envelope with and without disk models were 59 and 46$^{\circ}$ respectively, we therefore adopted an inclination of 50$^{\circ}$ to determine several of the physical properties of the jet and outflow in Section \ref{outflow_jet_VLA3}, as a compromise between these two values.

We found low cavity densities $\rho_{\rm cav}$ were preferred for all runs, with best-fitting values between 10$^{-21}$ and $\sim2\times10^{-20}$\,g\,cm${}^{-3}$. In \citet{wang12} and \citet{van-der-tak99}, their model cavity densities were set to zero.

Our modelling found the mass of the disk to be on the order of 10-20\,M$_\odot$, whereas using 0.5$''$ 203.4\,GHz dust continuum observations \citet{wang12} determined the disk mass to be $\sim$1-3\,M$_\odot$. However, as the disk mass was one of the most poorly determined parameters by our modelling, and as the interferometric dust continuum observations of \citet{wang12} directly measure the column density and thus mass of the disk, without the intervening emission or scattering from the envelope, we suggest that the result of \citet{wang12} is more robust.

It was not possible to compare our best-fitting values for the disk scale-height $h_{0}$ to the model of \citet{wang12}, as they instead used a $\sin^{5}{\theta}$ function to describe the vertical structure of the disk, where $\theta$ is the angle from the polar axis, as opposed to our Gaussian profile with z.

The best-fitting luminosity of AFGL\,2591 is 2.3$\times10^{5}$\,L$_{\odot}$ for both the envelope with and without disk models respectively, which is in agreement with previous results \citep[2.1-2.5$\times10^5\,$L$_{\odot}$ at 3.33\,kpc,][]{lada84,henning90,rygl12}.

The envelope density parameter $\rho_{\rm{env}0}~R_{\rm c}^{3/2}$ is a proxy for the mass of the envelope, which dominates the total mass of the enclosed circumstellar material at large radii, as it describes the scaling of equation (2). We find the total mass of the envelope plus disk is 1200\,M$_{\odot}$ for the envelope plus disk model, and 2700\,M$_{\odot}$ for the envelope without disk model. The total mass of gas associated with the region observed by \citet{minh08} is 2$\times$10$^{4}$\,M$_{\odot}$, which is scaled to 3.33\,kpc and within a radius of 5.8\,pc. Thus although our model traces a sizeable fraction of the envelope material within a radius of several pc, it may miss more diffuse material on the scales observed by \citet{minh08}, and thus could underestimate the true total mass associated with the region. To determine whether the mass distribution on smaller scales of our best-fitting models was consistent with other results, we compared the total mass within a radius of 2700\,AU to the model of \citet{jimenez-serra12}. Within this radius, which is half the determined size of their envelope 5400\,AU, the total mass of the model of \citet{jimenez-serra12} is 1\,M$_{\odot}$. Comparing this to our model we find an enclosed mass within 2700\,AU of 0.1\,M$_{\odot}$ for the model without a disk, and 1\,M$_{\odot}$ for the model with disk. Therefore in the case where there is a disk, the enclosed masses are in agreement.

\subsection{3.6\,cm and 7\,mm Continuum}
\label{obsresultcm}

\begin{figure}
\begin{center}
\includegraphics[width=8.8cm]{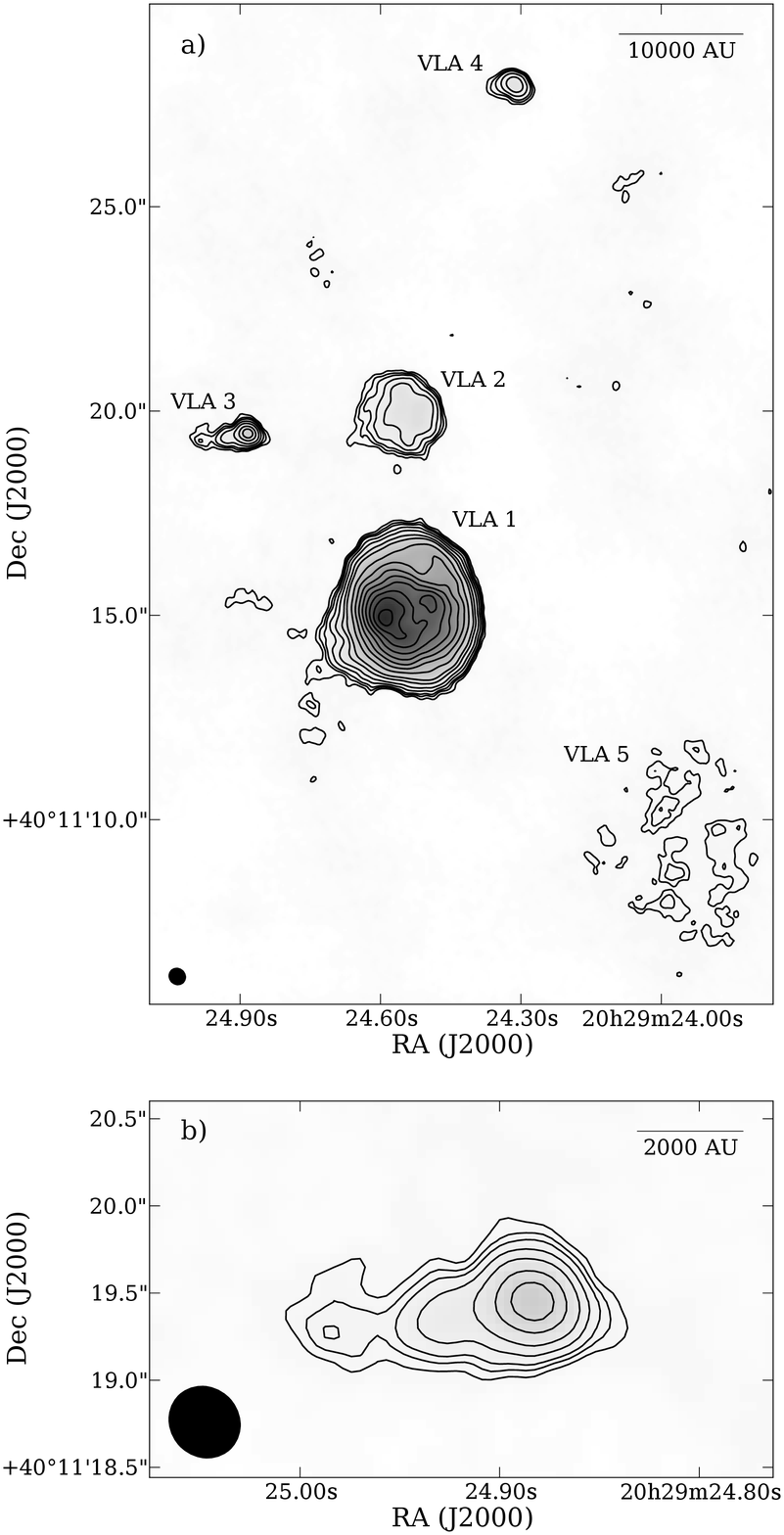}
 \caption[Map of the 3.6\,cm continuum emission surrounding AFGL\,2591]{a) Map of the 3.6\,cm continuum emission surrounding AFGL\,2591. Contours are -3, 3, 4, 5, 7, 10, 15, 20, 30, 40, 50... 100 $\times$ rms noise = 30$\mu$Jy\,beam$^{-1}$. Greyscale: -0.03 to 3.77 mJy\,beam$^{-1}$ (1.2 $\times$ peak value). The synthesised beam is shown in the bottom left corner: 0.43$''$ $\times$ 0.40$''$, P.A. = 43$^{\circ}$. b) Close-up of the 3.6\,cm continuum emission towards VLA\,3. Contours, greyscale and beam as in panel a).}
 \label{3.6cmfig}
 \end{center}
\end{figure}

\begin{figure}
\begin{center}
\includegraphics[width=8.8cm]{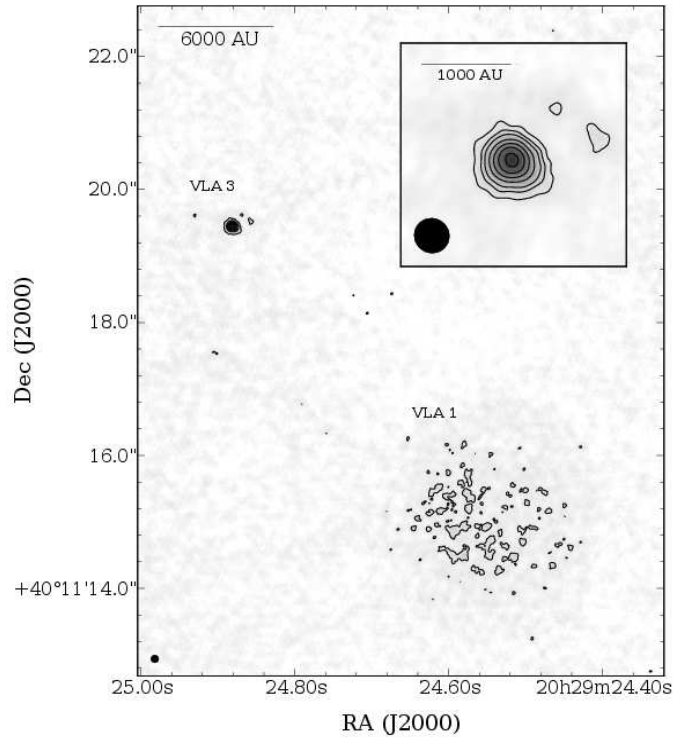}
 \caption[Map of the 7\,mm continuum emission towards AFGL\,2591.]{Map of the 7\,mm continuum emission towards AFGL\,2591. Contours are -4, 4, 7, 10, 15, 20, 25, 30 $\times$ rms noise = 56\,$\mu$Jy\,beam$^{-1}$. Greyscale: -0.06 to 2.14 mJy\,beam$^{-1}$ (1.2 $\times$ peak value). The inset panel shows a close-up of the 7\,mm emission from VLA\,3. The synthesised beam is shown in the bottom left corner of both images: 0.11$''$ $\times$ 0.11$''$, P.A. = 43$^{\circ}$.}
 \label{7mmfig}
 \end{center}
\end{figure}

\begin{table*}
 \centering
  \caption{Measured properties of the observed 3.6\,cm (8.4\,GHz) and 7\,mm (43\,GHz) continuum sources}  
\begin{tabular}{@{}lllllll@{}}
\hline
Source & $\lambda$ & Peak position & Peak flux & Integrated flux & Deconvolved & P.A. \\
name & (cm) & (J2000) & density (mJy\,beam$^{-1}$) & density (mJy) & source size ($''$) & (degrees) \\
\hline
VLA\,1 & 3.6 & 20 29 24.59  +40 11 15.0 & 3.14 $\pm$ 0.07 & 80 $\pm$ 1.6 & 4.9 $\times$ 4.0 & 120 \\
 & 0.7 & 20 29 24.580  +40 11 14.50 & 0.46 $\pm$ 0.06 & 64 $\pm$ 3.2 & 1.8 $\times$ 1.6 & 90\\
 VLA\,2 & 3.6 & 20 29 24.52 +40 11 20.1 & 0.45 $\pm$ 0.03  & 5.2 $\pm$ 0.11 & 2.5 $\times$ 2.3 & 110 \\
  & 0.7 & ... & ... & ... & ... & ... \\
 VLA\,3 & 3.6 & 20 29 24.88 +40 11 19.5 & 0.72 $\pm$ 0.03 & 1.52 $\pm$ 0.03 & 1.9 $\times$ 0.84 & 90 \\
 & 0.7 & 20 29 24.882 +40 11 19.45 & 1.8 $\pm$ 0.11 & 2.9 $\pm$ 0.14 & 0.35 $\times$ 0.26 & 50\\
 VLA\,4 & 3.6 & 20 29 24.32 +40 11 28.0 & 0.39 $\pm$ 0.03 & 0.99 $\pm$ 0.02 & 1.1 $\times$ 0.78 & 90\\
 & 0.7 & ... & ... & ... & ... & ... \\
 VLA\,5 & 3.6 & 20 29 24.00 +40 11 10.3 & 0.15 $\pm$ 0.03 & 10.9 $\pm$ 0.22 & 4.4 $\times$ 2.9 & 0\\
  & 0.7 & ... & ... & ... & ... & ... \\
\hline
\label{vlafluxtable}
\end{tabular}
\end{table*}

Figure \ref{3.6cmfig} presents the observed multi-configuration 3.6\,cm image of AFGL\,2591. Panel a) shows the entire region surrounding the source, including the four sources first observed by \citet{campbell840} at 6\,cm (VLA\,1 through 4). In addition, a new, low surface brightness source was detected: VLA\,5, which varies in peak flux from 3-5\,$\sigma$ across the extent of the source. With the addition of shorter baselines, more extended emission is recovered compared to the 3.6\,cm A array images of \citet{tofani95} and \citet{trinidad03}. Most noticeably, the emission from source VLA\,3, which is coincident with the position of the central source in the near-IR images, is better-determined; a close-up of the emission from this source is shown in panel b) of Fig.~\ref{3.6cmfig}. The 3.6\,cm emission from VLA\,3 is consistent with a compact core plus jet morphology. The dominant side of the jet extends to the east, with a deconvolved width and length of $<$0.2$''$ and 1.2$''$ ($<$670\,AU and 4000\,AU at 3.33\,kpc), position angle of $\sim$100$^{\circ}$, and an opening angle (at a radius of 4000\,AU) of $<$10$^{\circ}$ derived from its deconvolved width and length. The east jet is resolved, ending in a `knot' at 20$^{\rm h}$29$^{\rm m}$24$^{\rm s}$.98 +40$^{\circ}$11$'$19$''$.3 (J2000). There is also a smaller corresponding jet to the west, seen as a slight extension of the source in this direction. The morphology of VLA\,3 is consistent with a jet-system which is approximately parallel to the larger-scale flow \citep[observed by e.g.][]{hasegawa95}, and therefore these observations confirm the proposal of \citet{trinidad03} that VLA\,3 is its powering source.

Figure \ref{7mmfig} shows the 7\,mm multi-configuration image of the region surrounding AFGL\,2591, in which only VLA\,1 and VLA\,3 were detected at this resolution. VLA\,3 is compact, with a radius of $<500$\,AU, and is only slightly resolved in this image, showing no evidence of the eastern jet seen in Fig. \ref{3.6cmfig} at 3.6\,cm, which may partially be due to the 7\,mm images having almost twice the rms noise. However, the A array-only image of \citet{van-der-tak05}, which has a synthesised beam size of 43 by 37 mas, shows that the source is slightly extended to the south-west, in a similar direction to the counter-jet seen in the 3.6\,cm image. This extension is also seen in our multi-configuration 7\,mm image. Using VLBA water maser observations, \citet{sanna12} found that this emission most likely arises from the outflow cavity walls, which the masers they observed were also found to trace. 

Table \ref{vlafluxtable} provides the measured positions, peak and integrated fluxes, as well as deconvolved sizes for the sources observed in the 3.6\,cm (8.4\,GHz) and 7\,mm (43.3\,GHz) images. These were measured using a custom-made irregular aperture photometry program, with which the integrated flux density was measured above 1 sigma. Uncertainties in the aperture fluxes were calculated to be a combination of the uncertainty due to the image noise over the aperture, and the maximum VLA absolute flux error, which is 2\% and 5\% for 3.6\,cm and 7\,mm respectively. Similarly, the peak flux uncertainty was found by combining the 1 sigma flux density with the VLA absolute flux error. The deconvolved source size at the $3-\sigma$ level was measured by taking the major axis of the source to be along the direction which it is most extended, to within a position angle of 10$^{\circ}$.

\begin{figure}
\begin{center}
\sidecaption
\includegraphics[width=8.8cm]{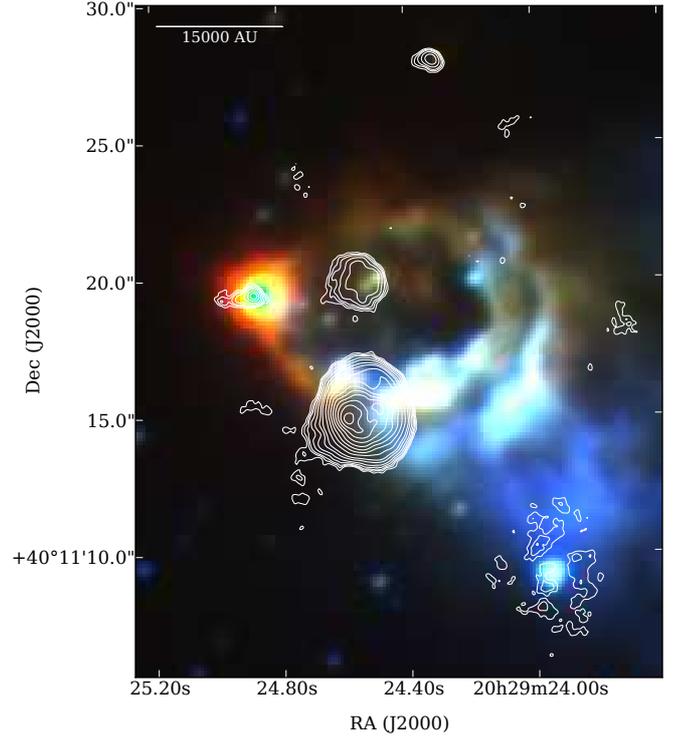}
\caption[Three-colour JHK$'$ Germini-North image of AFGL\,2591 overlaid with 3.6\,cm contours]{Three-colour JHK$'$ Germini-North image of AFGL\,2591 overlaid with the 3.6\,cm contours from Fig. \ref{3.6cmfig}. Stretch of Gemini image: red, K$'$ band: 150 - 2500 MJy\,sr$^{-1}$; green, H band: 150 - 500 MJy\,sr$^{-1}$; blue, J band: 40 - 80 MJy\,sr$^{-1}$.
}
\label{cmGemfig}
\end{center}
\end{figure}

Figure \ref{cmGemfig} compares the 3.6\,cm emission from the region to the near-IR emission recorded in the Gemini North images. The five detected H{\sc II} regions line-up with several features in the near-IR image. Firstly, the peak of VLA\,3 is coincident with that of the central source of AFGL\,2591, at the apex of the one sided reflection nebula. VLA\,1 is instead anti-correlated with the diffuse near-IR emission. This dent or cavity in the cloud can be seen more clearly in the K band bispectrum speckle interferometry image of \citet{preibisch03}. VLA\,2 also appears to be coincident with the close binary discovered by \citet{preibisch03}, and VLA\,4 with a source in the Gemini North image at 20$^{\rm h}$29$^{\rm m}$24$^{\rm s}$.3 +40$^{\circ}$11$'$28$''$ (J2000; however this source is too faint to be seen in Fig. \ref{cmGemfig}). In addition, VLA\,5 may be powered by the bright source at 20$^{\rm h}$29$^{\rm m}$23$^{\rm s}$.96 +40$^{\circ}$11$'$09$''$.25 (J2000; which is the same source as 2MASS 20292393+4011105, previously mentioned in Section \ref{pres2MASS}). The sources VLA\,1 and VLA\,3 are discussed in detail below. 

\subsubsection{The Central Source of AFGL\,2591, VLA\,3}
\label{vla3}

\begin{table}
 \centering
  \caption[Overview of observed fluxes for VLA\,1 to VLA\,5]{Overview of observed fluxes for VLA\,1 to VLA\,5. References are: (1) \citet{campbell840}, (2) this work (3) \citet{van-der-tak99}, (4) \citet{marengo00}, (5) \citet{trinidad03}}  
\begin{tabular}{@{}lllll@{}}
\hline
Source & Wavelength & Integrated flux & Ref. \\
name & & density (mJy) & & \\
\hline
VLA\,1 	& 6\,cm			& 79 $\pm$ 2.0 				& (1)	& 	\\ 
		& 3.6\,cm  		& 80 $\pm$ 1.6	 				& (2)	& 	\\ 
		& 7\,mm 		        & 64 $\pm$ 3.2 			        & (2)	& 	\\ 
		& 3.4\,mm 		& 87 $\pm$ 1.4 				& (3)	&     \\ 
		& 2.83\,mm		& 71 $\pm$ 1.2 				& (3)	&     \\ 
		& 2.81\,mm 		& 76 $\pm$ 15 (20\%)			& (2)	&     \\ 
		& 2.7\,mm         	& 72 $\pm$ 14 (20\%)			& (2)	&     \\ 
		& 2.6\,mm  		& 93 $\pm$ 2.5 				& (3)	&     \\ 
 		& 18.0\,$\mu$m	& 7.96 $\pm$ 0.8 $\times$ 10$^4$ 	& (4)	&     \\
		& 12.5\,$\mu$m	& 1.59 $\pm$ 0.2 $\times$ 10$^4$ 	& (4)	&     \\
		& 11.7\,$\mu$m	& 1.84 $\pm$ 0.2 $\times$ 10$^4$ 	& (4)	&    \\
		
 VLA\,2 	& 6\,cm			        & 3.61 $\pm$ 0.72 				& (1)	& 	\\ 
		& 3.6\,cm				& 5.2 $\pm$ 0.11	 			& (2)	& 	\\ 
				
  VLA\,3 	& 6\,cm				& 0.4 $\pm$ 0.1 					& (1) \\ 
 	 	& 3.6\,cm  			& 1.52 $\pm$ 0.03 					& (2)\\ 
 	 	& 1.3\,cm				& 1.57 $\pm$ 0.4					& (5)\\ 
 	 	& 7\,mm			 	& 2.9 $\pm$ 0.14 					& (2)\\ 
       		& 3.4\,mm			 	& 29.5 $\pm$ 0.8					& (3) \\ 
       		& 2.83\,mm			& 38.7 $\pm$ 0.7					& (3) \\ 
       		& 2.81\,mm	 	 	& 71 $\pm$ 14 (20\%)				& (2) \\ 
       		& 2.7\,mm			 	& 79 $\pm$ 16 (20\%)				& (2) \\ 
       		& 2.6\,mm				& 52.9 $\pm$ 1.5					& (3) \\ 
 	 	& 1.3\,mm				& $\sim$151 $\pm$ 4.5				& (3)\\ 
       		& 18.0\,$\mu$m 	& 7.53 $\pm$ 0.75 $\times$ 10$^5$		& (4) \\
       		& 12.5\,$\mu$m 	& 7.45 $\pm$ 0.75 $\times$ 10$^5$		& (4) \\
       		& 11.7\,$\mu$m 	& 4.40 $\pm$ 0.44 $\times$ 10$^5$		& (4)\\
		
 VLA\,4 	& 6\,cm			        & 0.65 $\pm$ 0.13 				& (1)	& 	\\ 
		& 3.6\,cm		                & 0.99 $\pm$ 0.02	 			& (2)	& 	\\ 
	
 VLA\,5 	& 3.6\,cm		 		& 10.9 $\pm$ 0.22	 			& (2)	& 	\\	
		
\hline
\label{othersfluxtable}
\end{tabular}
\end{table}

\begin{figure}
\begin{center}
\includegraphics[width=8.8cm]{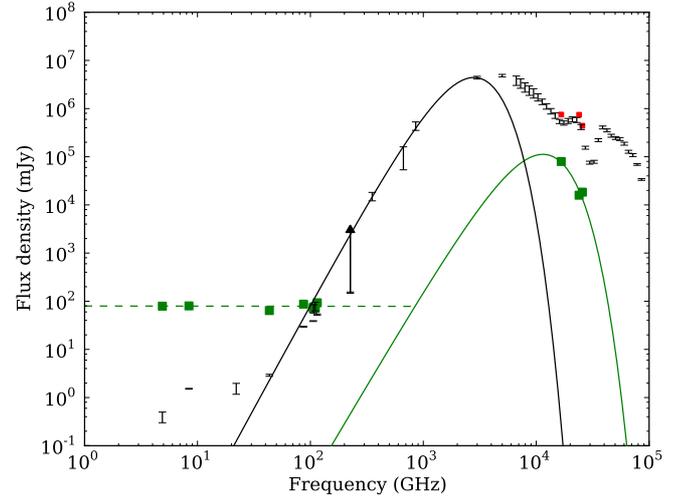}
\caption[Radio-SEDs of AFGL\,2591-VLA\,3 and VLA\,1]{Radio-SEDs of AFGL\,2591-VLA\,3 and VLA\,1. The SED of AFGL\,2591-VLA\,3 is show in black error bars, and the resolved fluxes of AFGL\,2591-VLA\,3 from \citet{marengo00} are shown as red squares. Green squares show the fluxes of VLA\,1 - errors are not shown as they are smaller than the markers. The black and green solid lines show greybody fits to the SEDs of AFGL\,2591-VLA\,3 between 100 and 850\,$\mu$m, and VLA\,1 at mid-IR fluxes, respectively (see Sections \ref{vla3} and \ref{vla1}). The green dashed line shows the fit to the long-wavelength fluxes for VLA\,1.}
\label{cmSEDfig}
\end{center}
\end{figure}

Figure \ref{cmSEDfig} shows the SED in frequency space of the central source of AFGL\,2591, VLA\,3, in black error bars. The fluxes shown for VLA\,3, which are taken from the literature and this work, are listed in Tables \ref{obstable} (for $\lambda<$1\,mm) and \ref{othersfluxtable} (for $\lambda>$1\,mm). At $\lambda>$1\,mm, fluxes have been listed from the observations with the largest available \emph{uv} coverage.
 
A greybody was fitted to the fluxes between 100 and 850\,$\mu$m, giving a temperature of 25\,$\pm$7\,K and a dust emissivity exponent $\beta$ of 2.3$\pm$0.6. For these four fitted fluxes, the image photometry apertures contained most if not all of the source flux.

It can be seen in Fig. \ref{cmSEDfig} that the fluxes at millimetre wavelengths (87-226\,GHz) fall short of the flux expected from the fitted greybody. However, as these fluxes are taken from interferometric observations, it is likely that they only contribute a fraction of the actual flux. For instance, \citet{van-der-tak05} note that their quoted 226\,GHz continuum flux represents only 5\% of the total flux, due to lack of shorter baselines. An arrow is shown on Fig. \ref{cmSEDfig} representing this correction, which agrees well with the flux expected at 226\,GHz from the fitted greybody.

Extrapolating this greybody to 7\,mm, a flux of 2.3\,mJy is expected, but the measured flux of VLA\,3 at 7\,mm is 2.9\,mJy. However, it is likely that a portion of the 7\,mm flux is due to ionized gas emission as well as dust emission, and a significant fraction of this combined emission is resolved out by the interferometer. To estimate the contribution from dust emission at 7\,mm, a model image was made of the best-fitting envelope with disk model found from the SED and image modelling presented in Section \ref{sedresults_afgl}. The 7\,mm VLA A to D array observations were simulated from the model image using the CASA task simdata, then combined and imaged. The resultant 7\,mm model image did not have any significant emission above 3$\sigma$ = 0.17\,mJy\,beam$^{-1}$ over the 0.1\,arcsec$^2$ photometry aperture for VLA\,3, giving an upper limit of 0.57\,mJy integrated over the aperture. Hence, we estimate that $>$2.3\,mJy of the observed 7\,mm emission is from ionized gas, and $<$0.57\,mJy is due to dust emission. This is in agreement with the results of \citet{sanna12} who found that the morphology of the 7\,mm emission seen in previous VLA A array observations most likely arises from photoionization of the outflow cavity walls.

To measure the spectral index ${\alpha}$ (where $S_{\nu} \propto \nu^{\alpha}$) of VLA\,3, which may give further insight into the nature of the emission, the 3.6\,cm and 7\,mm data were re-imaged using \emph{uv} distances which were common to both datasets, from 4.5 to 1037\,k$\lambda$, and the fluxes remeasured. The integrated fluxes were found to be 1.43$\pm$0.03\,mJy for 3.6\,cm and 3.3$\pm$0.17\,mJy for 7\,mm (which increased compared to the original 7\,mm flux due to a reduction in resolution, so that fainter emission was brought above the noise level). Therefore the spectral index between 3.6\,cm and 7\,mm was found to be 0.5$\pm$0.02, where the quoted error in the spectral index is due solely to the photometric errors, similar to the spectral index expected for an ionized wind, 0.6. However, as this value is measured from only two fluxes, it is likely to be more uncertain than the photometric errors given. This spectral index is also an upper limit, as we estimate a fraction of the 7\,mm flux is due to dust emission. By applying the common \emph{uv} distance range above, we derive $<$0.51\,mJy for the contribution from dust emission, and therefore $>$2.8\,mJy from ionized gas emission. Using this lower limit for the ionized gas emission at 7\,mm, we estimate the spectral index to be between 0.4~and~0.5.

\begin{figure*}[!htbp]
\begin{center}
\sidecaption
\includegraphics[width=12cm]{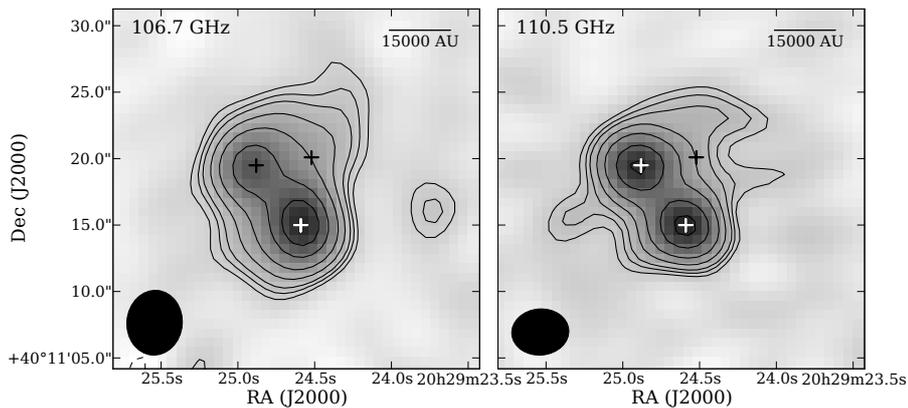}
 \caption[Continuum emission towards AFGL\,2591 at 106.7 and 110.5\,GHz ($\sim$2.8 and 2.7\,mm respectively)]{Continuum emission towards AFGL\,2591 at 106.7 and 110.5\,GHz ($\sim$2.8 and 2.7\,mm respectively), shown in both contours and greyscale. The rms noise in the images is $\sigma=$2.2 and 2.1\,mJy\,beam$^{-1}$ respectively. The greyscales extend from $-3\times \sigma$ to 65.5 and 54.0\,mJy\,beam$^{-1}$. Contours are at $-3,3,4,5,7,10,15 \rm{~and~} 20 \times \sigma$. The crosses show the positions of VLA\,1 to 3. The synthesised beams are 4.9$'' \times$4.1$''$ P.A.$=-4.8^{\circ}$ and 4.3$'' \times$ 3.5$''$ P.A.$=95^{\circ}$ respectively.}
 \label{3mm_cont}
 \end{center}
\end{figure*} 

\subsubsection{VLA\,1}
\label{vla1}
Figure \ref{cmSEDfig} shows the SED of VLA\,1 as green squares. This source was detected in the mid-IR by \citet{marengo00}, as well as at several radio frequencies, for which the fluxes are listed in Table \ref{othersfluxtable}. The spectrum of VLA\,1 appears to be flat at radio wavelengths, with a fitted spectral index $\alpha$ of $0.0\pm0.03$. The fitted power-law is shown as a dashed line. However, we first need to verify whether only a fraction of the flux of VLA\,1 is being recovered at each wavelength. The flux of VLA\,1 measured in our 3.6\,cm A-D array image is not larger than the fluxes measured in the A array only images in previous works (A-D array: 80\,mJy, this work; A array only: 82\,mJy, 94\,mJy, \citet{tofani95} and \citet{trinidad03} respectively). In fact, there is a small decrease in the flux of VLA\,1 for the larger \emph{uv} coverage data. Therefore it is likely that most of the flux from VLA\,1 has been recovered by observations which are sensitive to scales up to 7$''$, the largest observable angular scale of VLA A array observations at 3.6\,cm. This is the case for all of the interferometric fluxes shown for VLA\,1, hence the calculated flat spectral index of VLA\,1 is not likely to be due to instrumental effects. The observed spectral index is close to that expected from optically thin free-free ionized gas emission ($\alpha=-0.1$). 

The mid-IR emission from VLA\,1 is likely to arise from thermal emission from dust within the H{\sc II} region. To estimate how much VLA\,1 contributes to the total flux of the region, a greybody spectrum with a dust opacity index $\beta=2$ was fit to the fluxes measured by \citet{marengo00}. The best-fitting greybody has a temperature of 111$\pm$14\,K and peaks at 26$\pm$3.3\,$\mu$m with a flux of 112\,Jy, which is uncertain by a factor of two. Therefore the thermal dust emission from VLA\,1 never contributes more than $\sim$30\% to the total flux from the region at wavelengths between 8 and 100\,$\mu$m and a negligible fraction of the observed flux ($<0.5$\%) at other wavelengths, and therefore should not significantly affect the SED and image modelling results given in Section \ref{sedresults_afgl}. The mid-IR fluxes for only the resolved central source of AFGL\,2591 from \citet{marengo00} are plotted on Fig. \ref{cmSEDfig} (red squares) and agree well with the ISO-SWS data, which were used to cover the mid-IR to far-IR section of the SED. 

\subsection{$^{13}$CO, C$^{18}$O and 3\,mm Continuum}
\label{obsresultmm}
This section details the properties of the dense molecular gas surrounding AFGL\,2591 inferred from the C$^{18}$O molecular line and millimetre continuum observations. 

Figure \ref{3mm_cont} shows the continuum emission from the region at 106.7 and 110.5\,GHz, or 2.81 and 2.7\,mm, in both of the wide $\sim$3\,mm continuum bands observed. The morphology of the emission is similar to that seen in the millimetre continuum maps at various wavelengths presented by \citet{van-der-tak99}, \citet{van-der-tak06}, and resolved further by \citet{wang12}. We found that the sources in the 3\,mm images were systematically offset from the 3.6\,cm peak positions, which we determined to be caused by inaccurate coordinates used for the phase calibrator MWC\,349 in the mm observations. We therefore shifted the continuum and line maps by 0.89$''$ in R.A. and 0.18$''$ in declination to agree with the 7\,mm position of MWC\,349\,A reported in \citet{rodriguez07}. In Fig. \ref{3mm_cont}, the peak positions of VLA\,1 to 3 are shown as crosses. The main two sources that are detected are VLA\,1 and VLA\,3, but there is also extended emission towards the position of VLA\,2. Although the large uncertainties in the two measured 3\,mm fluxes do not allow calculation of accurate spectral indices for VLA\,1 and 2 using these values alone, the difference in the spectral index of the two sources can be seen from the difference in their relative brightnesses between the two images, indicating VLA\,1 has a flat spectral index, and VLA\,3 has a rising spectral index at shorter wavelengths. In addition, when we compared both images convolved with the beam of the 106.7\,GHz image, the morphology of the emission was very similar, apart from the clear difference in the flux of VLA\,3. The sources were fit with 2D Gaussians using the CASA task imfit to find the fluxes, which are given in Table~\ref{othersfluxtable}.

\begin{figure}[!htbp]
\begin{center}
\includegraphics[width=8.8cm]{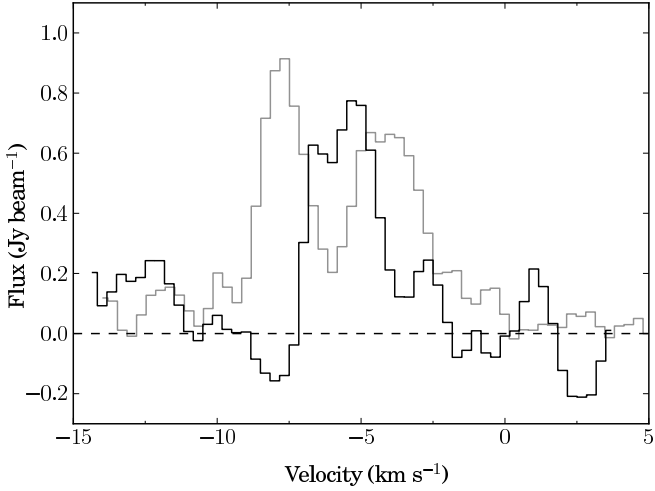}
\caption[C$^{18}$O and $^{13}$CO spectral profiles measured at the position of the central source of AFGL\,2591]{C$^{18}$O (black line) and $^{13}$CO (grey line) spectral profiles measured at the position of AFGL\,2591-VLA\,3: 20$^{\rm h}$29$^{\rm m}$24$^{\rm s}$.88 +40$^{\circ}$11$'$19$''$.5 (J2000).}
 \label{C18O_profile}
 \end{center}
\end{figure}

Figure \ref{C18O_profile} presents C$^{18}$O and $^{13}$CO line profiles at the position of AFGL\,2591-VLA\,3. The $^{13}$CO observations were missing significant flux on extended scales which, in combination with self-absorption effects, is likely to be the cause of the dip in the $^{13}$CO line profile seen in Fig. \ref{C18O_profile}. As there was a significant amount of missing flux in the $^{13}$CO line, and the emission was distributed incoherently across the maps, it was not possible to interpret this emission. Therefore the $^{13}$CO data will not be discussed further, however channel maps of the $^{13}$CO emission are included in the Online Material as Figs. \ref{13CO_channel} and \ref{13CO_channel2}.

Figure \ref{C18O_channel} presents channel maps of the C$^{18}$O emission towards AFGL\,2591 between -9.0 and -2.7\,km\,s$^{-1}$ (where {-5.7\,km\,s$^{-1}$} is the rest velocity of the cloud), and Fig. \ref{C18O_mom1} shows the C$^{18}$O intensity-weighted first moment map. In both Figures the positions of VLA\,1 through 3 are marked by crosses, however in Fig. \ref{C18O_mom1}, the position and size of VLA\,1 at 3.6\,cm is shown instead as a circle. A blue-shifted velocity feature can be seen in Fig. \ref{C18O_mom1} to the west of VLA\,3, and coincident with VLA\,2, which decreases smoothly in velocity to the west VLA\,3, ranging from -5.3 to -8.0\,km\,s$^{-1}$. This can be seen in the C$^{18}$O channel maps (Fig. \ref{C18O_channel}), which also show that the shape of the emission becomes narrower away from the line centre and towards the west. This is most obvious in the -7.7 and -8.0\,km\,s$^{-1}$ channels. In addition, Fig. \ref{C18O_mom1} shows that there is a smaller intensity peak $\sim$6$''$ to the south of VLA\,3. Here, the velocity instead decreases from northwest to southeast from approximately -5 to -6\,km\,s$^{-1}$.

In Fig. \ref{Gemini_compare}, integrated C$^{18}$O emission in several velocity ranges is compared with the Near-IR Gemini image, and the positions of bow shocks detected in the K$'$ band Gemini North image \citep[first noted by][]{preibisch03}. The bow shocks are shown approximately as white semi-ellipses, as they are too faint to be seen directly from Fig. \ref{Gemini_compare}. The white dashed line also shows the direction of the ionized jet seen in the 3.6\,cm images (Fig. \ref{3.6cmfig}), with P.A. 100$^{\circ}$. The contours show the high-velocity red and blue-shifted gas (red and blue contours respectively). Although the ``high"-velocity channels referred to in this work are high velocity relative to the observed line, it should be noted that more extended, higher velocity gas has been detected by \citet{hasegawa95} using single dish $^{12}$CO observations with 14.3$''$ resolution, which extends in velocity from -45 to 35\,km\,s$^{-1}$. Therefore, the observations presented here trace the higher density but comparatively lower-velocity gas, within a few km\,s$^{-1}$ of the line centre. To minimise confusion due to overlap in velocity of various components, the contours in Fig. \ref{Gemini_compare} only show integrated C$^{18}$O emission from -4.3 to -3.3\,km\,s$^{-1}$ (red) and -8.0 to -7.0\,km\,s$^{-1}$ (blue). 

\onlfig{11}{
\begin{figure*}
\begin{center}
\includegraphics[width=18cm]{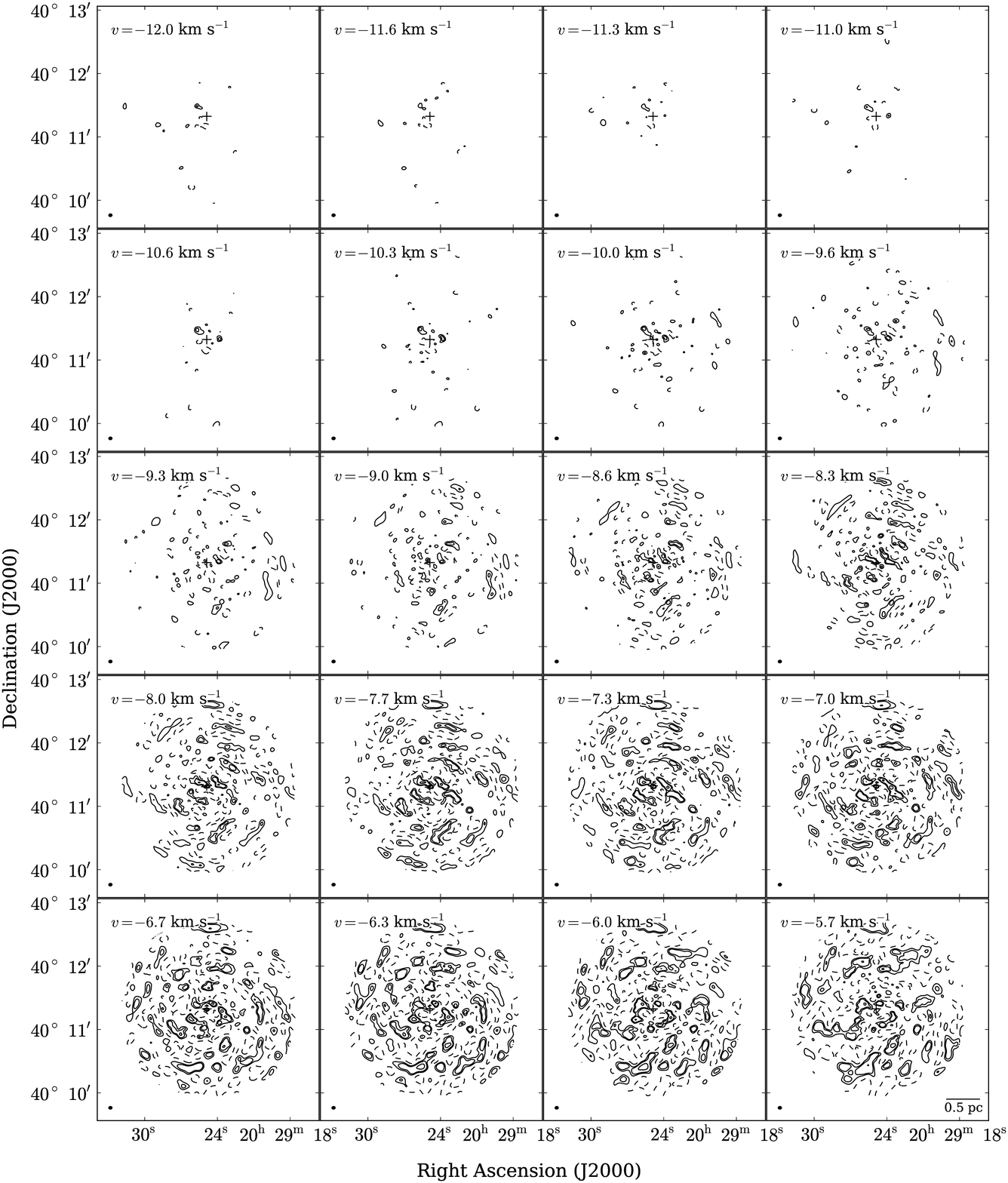}
 \caption[$^{13}$CO channel map at 0.3\,km\,s$^{-1}$ resolution between -12.0 and -5.7\,km\,s$^{-1}$]{$^{13}$CO channel map at 0.3\,km\,s$^{-1}$ resolution between -12.0 and -5.7\,km\,s$^{-1}$. The rms in the map is $\sigma$=0.1\,Jy\,beam$^{-1}$. The map peak flux is 1.3\,Jy\,beam$^{-1}$. Contours are at -3, 3, 5, 10, 15 $\times\,\sigma$. The synthesised beam is shown in the bottom left-hand corner (4.4$'' \times$ 3.7$''$, P.A. 96$^{\circ}$). A scale size of 0.5\,pc is represented by a bar in the bottom right-hand panel. The cross shows the position of the central source of AFGL\,2591 from the Gemini near-IR J band image. The rest velocity of the cloud is -5.7\,km\,s$^{-1}$.}
 \label{13CO_channel}
 \end{center}
\end{figure*}
}

\onlfig{12}{
\begin{figure*}
\begin{center}
\includegraphics[width=18cm]{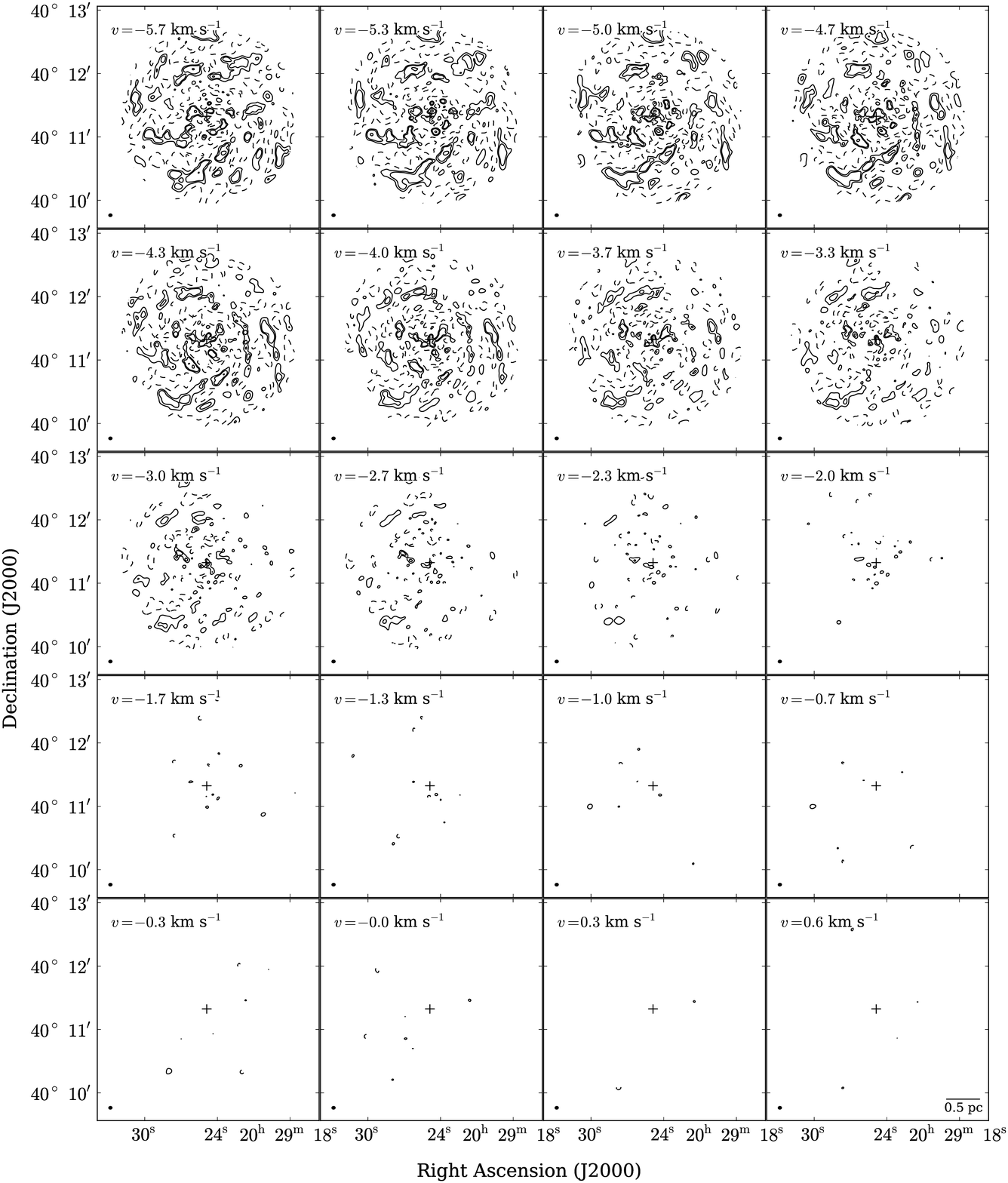}
 \caption[$^{13}$CO channel map at 0.3\,km\,s$^{-1}$ resolution between -5.7 and 0.6\,km\,s$^{-1}$]{$^{13}$CO channel map at 0.3\,km\,s$^{-1}$ resolution between -5.7 and 0.6\,km\,s$^{-1}$. The rms in the map is $\sigma$=0.1\,Jy\,beam$^{-1}$. The map peak flux is 1.3\,Jy\,beam$^{-1}$. Contours are at -3, 3, 5, 10, 15 $\times\,\sigma$. The synthesised beam is shown in the bottom left-hand corner (4.4$'' \times$ 3.7$''$, P.A. 96$^{\circ}$). A scale size of 0.5\,pc is represented by a bar in the bottom right-hand panel. The cross shows the position of the central source of AFGL\,2591 from the Gemini near-IR J band image. The rest velocity of the cloud is -5.7\,km\,s$^{-1}$.}
 \label{13CO_channel2}
 \end{center}
\end{figure*}
}

\section{Discussion}
\label{discussion}

\subsection{The Jet and Outflow from VLA\,3}
\label{outflow_jet_VLA3}

In Fig. \ref{Gemini_compare}, the extrapolated direction of the ionized jet observed at 3.6\,cm points directly through the red-shifted C$^{18}$O contours to the bow-shocks seen in the K$'$ band image. Hence this outlines a coherent picture in which the red-shifted outflow lobe of AFGL\,2591 consists of a small-scale 4000\,AU ionized jet that is part of a jet or wind extending out to $\sim$0.4\,pc where the flow terminates against the surrounding cloud as bow-shocks. However it is interesting to note that the position angle of the jet and bow shocks do not exactly align with that of the blue-shifted reflection nebula or the elongation of the blue-shifted outflow observed by \citet{hasegawa95}. Therefore other stars in the vicinity, such as the powering star(s) of VLA\,1, may be causing precession of the jet. If, as well as the emission lying along the position angle of the ionized jet, the clump of emission at 20$^{\rm h}$29$^{\rm m}$26$^{\rm s}$.7 +40$^{\circ}$11$'$23$''$ (J2000) is also included, these three clumps may be tracing an arc of emission describing the densest parts of the red-shifted outflow lobe, created as the jet precesses. If so, the extent of these clumps corresponds to an observed opening angle of $\sim$40$^{\circ}$ at their distance from the source, $\sim$0.4\,pc. At the same distance, the observed blue-shifted outflow lobe is larger with an opening angle of $\sim$60$^{\circ}$. The opening angles of the biconical outflow observed by \citet{jimenez-serra12} and the blue shifted outflow cone observed by \citet{sanna12}, on scales $<$2$''$ or $<$6700\,AU, are $\sim$90-110$^{\circ}$. The larger observed opening angles on smaller scales suggests that the outflow cavity can not be described by a cone with a constant opening angle, but is instead better described by a power-law cavity similar to that used in our radiative transfer models.

Several suggestions regarding the nature of the mm and cm continuum emission from VLA\,3 have been made by previous studies, including a core-halo H{\sc II} region, an ionized wind with a dust disk \citep{trinidad03}, or emission from a spherical gravitationally confined H{\sc II} region \citep{van-der-tak05}. At 3.6\,cm, the deeper observations presented here clearly show that as well as a central compact core, the source also exhibits a non-spherical jet-like morphology. Therefore, we calculate several properties of the emission below, assuming it originates from a jet.

Without assuming a specific ionization mechanism, the mass loss rate of the jet can be estimated using the model of \citet{reynolds86}, which describes the emission from a partially optically thick ionized jet (his equation 19):
\begin{equation}
\begin{split}
\frac{\dot{M}}{10^{-6}\,\rm{M}_{\odot} \rm{yr}^{-1}} = 9.38\times10^{-2} \left( \frac{\upsilon}{100\,\rm{km\,s}^{-1}} \right) \left( \frac{1}{x_0} \right) \left( \frac{\mu}{m_p} \right) \\
\times ~~ \left[ \left( \frac{S_{\nu}}{\rm{mJy}} \right) \left( \frac{\nu}{10\,\rm{GHz}} \right)^{-\alpha} \right]^{0.75} \left( \frac{d}{kpc} \right)^{1.5} \left( \frac{\nu_{m}}{10\,\rm{GHz}} \right)^{-0.45+0.75\alpha} \\
\times ~~ \theta^{0.75} \left(\frac{T}{10^4\,\rm{K}}\right)^{-0.075} (\sin{i})^{-0.25}~F^{-0.75}
\end{split}
\end{equation}
where $\upsilon$ is the velocity of the jet, $x_0$ is the ionization fraction; $\mu/m_p$ is the mean particle mass per hydrogen atom of the ionized material, given by $1/(1+x_0)$; $S_{\nu}$ is the observed flux at the frequency $\nu$; $\alpha$ is the spectral index; $d$ is the distance to the source; $\nu_m$ is the turn-over frequency below which the emission becomes optically thick; $\theta$ is the opening angle of the flow, defined as the ratio of the projected width to the radius at the base of the jet; $T$ is the temperature of the ionized gas; $i$ is the inclination measured from the line-of-sight and $F$ is a function of the spectral index and the dependance of optical depth on radius \citep[see][equation 17]{reynolds86}.

Assuming an isothermal, uniformly ionized jet with a density gradient of $\rho\sim r^{-2}$; $\upsilon=500$\,km\,s$^{-1}$ for the velocity of the jet, found for the blue-shifted HH objects towards AFGL\,2591 \citep{poetzel92}; $x_0=0.1$ \citep[commonly found for low-mass sources, e.g.][]{bacciotti99}; $v_m=50$\,GHz; $\theta\sim$ 0.2$''/$1.2$''$, the ratio of the maximum deconvolved width to the maximum radius; $T=10^4$\,K ; a spectral index between 0.4 and 0.5; and $i=50^{\circ}$, taken from the results of the SED and near-IR image profile modelling presented in Section \ref{sedresults_afgl}; $F$ was found to be between 2.3 and 1.8, and the mass loss rate of the jet observed at 3.6\,cm was therefore determined to be in the range 0.77 - 1.0$\times10^{-5}\,\rm{M}_{\odot} \rm{yr}^{-1}$.

\begin{figure*}[!htbp]
\begin{center}
\includegraphics[width=18cm]{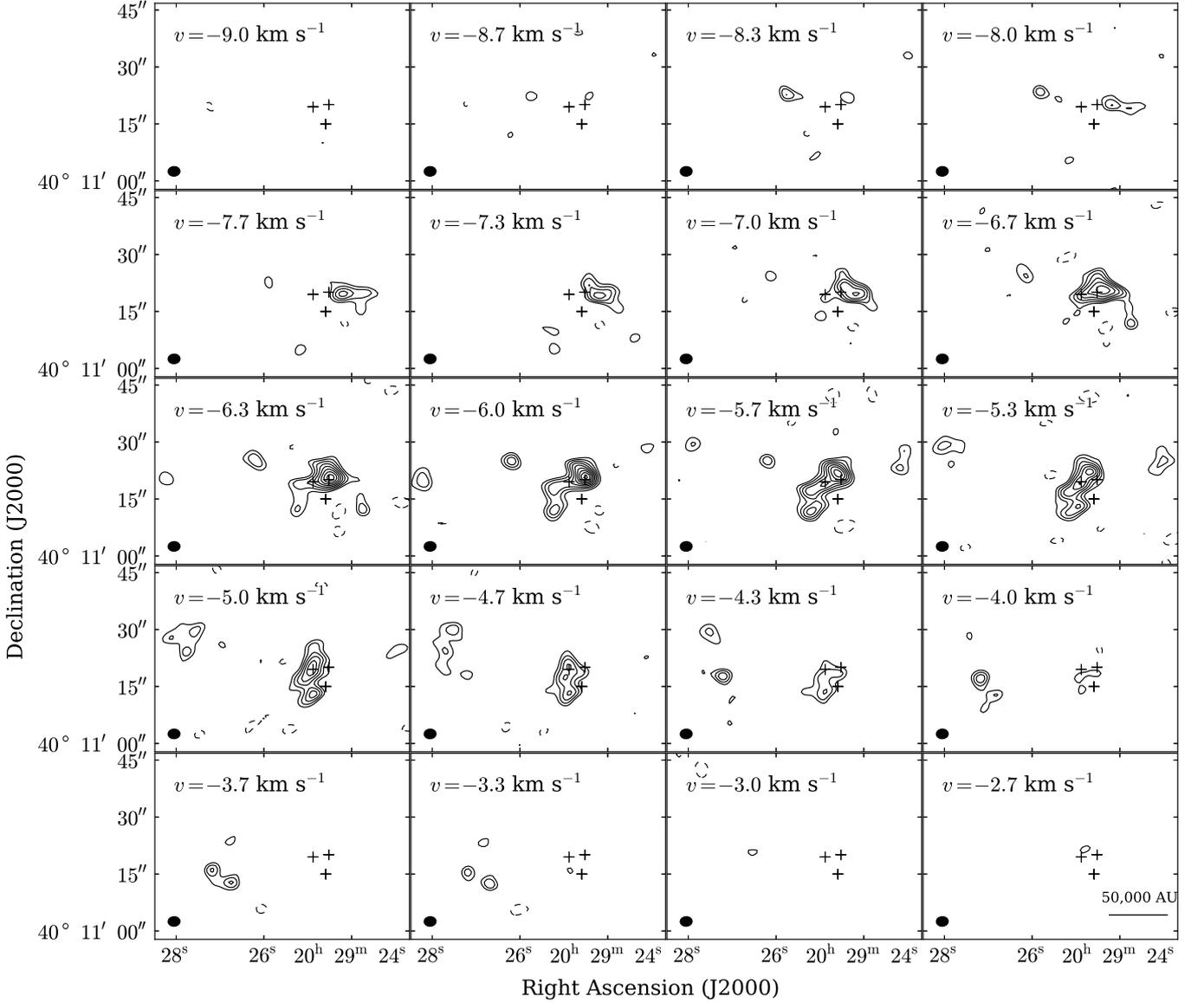}
 \caption[C$^{18}$O channel map at 0.3\,km\,s$^{-1}$ resolution between -9.0 and -2.7\,km\,s$^{-1}$]{C$^{18}$O channel map at 0.3\,km\,s$^{-1}$ resolution between -9.0 and -2.7\,km\,s$^{-1}$. The rms in the map is $\sigma$=0.1\,Jy\,beam$^{-1}$. The map peak flux is 1.1\,Jy\,beam$^{-1}$. Contours are at -3, 3, 4, 5, 6, 7, 8, 9, 10, 11 $\times\,\sigma$. The synthesised beam is shown in the bottom left-hand corner (4.5$'' \times$ 3.6$''$, P.A. 93$^{\circ}$). A scale size of 50,000\,AU is represented by a bar in the bottom right-hand panel. The cross shows the position of the central source of AFGL\,2591 from the Gemini near-IR J band image. The rest velocity of the cloud is -5.7\,km\,s$^{-1}$.}
 \label{C18O_channel}
 \end{center}
\end{figure*}

This value is approximately a thousand times larger than the jet mass loss rates seen for low-mass protostars, which are commonly found to be on the order of 10$^{-8}\,\rm{M}_{\odot} \rm{yr}^{-1}$ \citep[e.g.][]{podio06}. The mass loss rate in this jet can also be compared to the mass loss rate of the small-scale red-shifted molecular flow observed by \citet{hasegawa95} in $^{12}$CO(J=3-2), which corresponds well to the position and direction of the outflow lobe suggested by the ionized jet, red-shifted C$^{18}$O emission and bow shocks discussed in Section \ref{obsresultmm}. For their optically thick case, \citeauthor{hasegawa95} find a mass loss rate of $6.3\times10^{-5}$\,M$_{\odot}$yr$^{-1}$ at 1\,kpc corresponding to $7.0\times10^{-4}$\,M$_{\odot}$yr$^{-1}$ at 3.33\,kpc. 

Multiplying the ionized jet mass loss rate by the velocity $\upsilon$, the momentum rate in the jet is 3.9 - 5.2$\times10^{-3}$\,M$_{\odot}$yr$^{-1}$km\,s$^{-1}$, which is very similar to the momentum rates determined by \citet{hasegawa95} for both the large and small scale red-shifted $^{12}$CO outflows (7.7 and 7.4$\times10^{-3}$\,M$_{\odot}$yr$^{-1}$km\,s$^{-1}$ at 3.33\,kpc with $i=50^{\circ}$, scaled from 7.5 and 7.2$\times10^{-4}$\,M$_{\odot}$yr$^{-1}$km\,s$^{-1}$ at 1\,kpc with $i=45^{\circ}$). Thus the jet itself would have very close to the required momentum to drive the observed larger scale red-shifted outflow.

The ionization mechanism of the jet can be modelled as a plane-parallel shock in a homogeneous neutral `stellar' wind \citep[e.g.][]{curiel89,anglada96}. The momentum rate $\dot{P}$ of the jet can be expressed as:
\begin{equation}
\begin{split}
\left( \frac{\dot{P}}{\rm{M}_{\odot}\,\rm{yr}^{-1}\,\rm{km\,s}^{-1}} \right) =
 \frac{3.13\times10^{-4}}{\eta} \left( \frac{S_{\nu}d^{2}}{\rm{mJy}\,\rm{kpc}^2} \right) \left( \frac{\upsilon_{\star}}{200\,\rm{km\,s}^{-1}} \right)^{0.32} 
\\
\times ~~ \left( \frac{T}{10^4\,\rm{K}} \right)^{-0.45} \left( \frac{\nu}{5\,{\rm GHz}} \right)^{0.1} \left( \frac{\tau}{1-e^{-\tau}}\right)
\end{split}
 \end{equation}
where $\eta$ is the shock efficiency or ionization fraction, found to be $\sim$0.1 for low-mass sources \citep[e.g.][]{bacciotti99}, $\upsilon_{\star}$ is the initial velocity of the stellar wind or jet, taken to be 500\,km\,s$^{-1}$, and $\tau$ is the optical depth of the emitting gas. Here, the flux $S_{\nu}$ is measured at 3.6\,cm. The optical depth can be determined using equation 6 of \citet{anglada98}:
\begin{equation}
\alpha = 2 + \frac{\ln{[(1-e^{-\tau_1})/(1-e^{-\tau_1(\nu_1/\nu_2)^{2.1}})]}}{\ln{(\nu_1/\nu_2)}}
\end{equation}
Using the estimated spectral index of 0.4 - 0.5, the optical depth of the emission at 3.6\,cm was estimated to be between 2.1 and 2.6, giving a value for the required momentum rate in the jet of 2.8 - 3.4$\times10^{-2}$\,M$_{\odot}$yr$^{-1}$km\,s$^{-1}$. Therefore the required momentum rate in the jet is larger than the value of 3.9 - 5.2$\times10^{-3}$\,M$_{\odot}$yr$^{-1}$km\,s$^{-1}$ calculated above by a factor of $\sim$5 to 9, implying that the 3.6\,cm emission cannot be solely caused by shocks in the jet. However, if the shock efficiency is increased to near unity, the momentum rate of the jet would be sufficient to ionize it. As this is unlikely, photoionization by the central star therefore must provide a significant fraction of the emission at 3.6\,cm.

\begin{figure}[!htbp]
\begin{center}
\includegraphics[width=8.8cm]{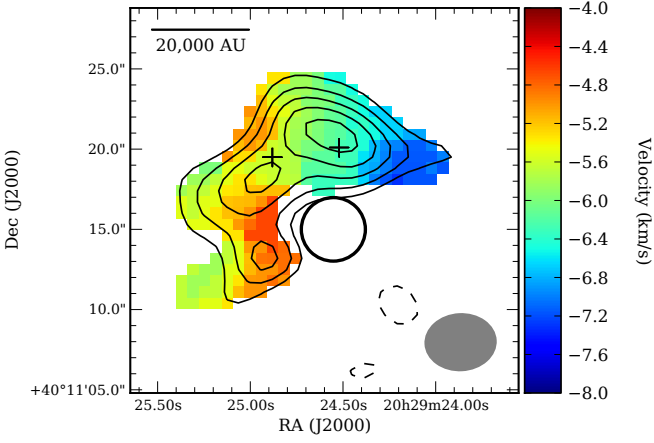}
 \caption[Intensity-weighted C$^{18}$O first moment map overlaid with C$^{18}$O integrated intensity contours]{Shown in colour-scale: intensity-weighted first moment map of the C$^{18}$O emission towards AFGL\,2591. The corresponding velocities are shown on the colour-bar on the right. The black open circle represents the position and size of VLA\,1 at 3.6\,cm. The crosses indicate the position of VLA\,2 and VLA\,3. A scale size of 20,000 AU is represented by a bar in the upper left-hand corner. Shown in contours: the C$^{18}$O integrated intensity map from -2.3 to -8.7\,km\,s$^{-1}$. The peak flux of the integrated intensity map is 2.5\,Jy\,beam$^{-1}$\,km\,s$^{-1}$. Contours are at -3, 3, 4, 5, 6, 7 $\times\,\sigma$=0.33\,Jy\,beam$^{-1}$\,km\,s$^{-1}$. The synthesised beam is shown in the lower right-hand corner (4.5$'' \times$ 3.6$''$, P.A. 93$^{\circ}$). The rest velocity of the cloud is -5.7\,km\,s$^{-1}$.}
 \label{C18O_mom1}
 \end{center}
\end{figure}

Turning to the C$^{18}$O observations, the blue-shifted emission seen in Fig.\ref{C18O_mom1} and the blue contours in Fig. \ref{Gemini_compare} may be tracing a blue-shifted outflow from VLA\,2 or VLA\,3. If this is the case, the outflow can be seen to be narrower and more collimated at more negative velocities, seen in Fig. \ref{C18O_channel}. In addition, a ``Hubble-like" velocity trend \citep[i.e. an systematic increase flow velocity with distance from the source, e.g.][]{lada96} can be seen across the blue-shifted emission in Fig. \ref{C18O_mom1}. Therefore there is a high-velocity collimated component at the centre and front of this blue-shifted flow, which smoothly transitions into a surrounding lower-velocity wide-angle wind.

Although the integrated C$^{18}$O emission peaks on VLA\,2, thus making it a possibility that VLA\,2 is driving the outflow, there are several lines of evidence which suggest this is not the case. Firstly, the H{\sc II} region surrounding VLA\,2 appears to be more evolved than VLA\,3, and hence VLA\,2 is less likely to be driving an outflow. Secondly, the emission at the position of VLA 3 is at the rest velocity of the cloud (5.7\,km\,s$^{-1}$) instead of VLA 2, where it is approximately 6.3\,km\,s$^{-1}$. Thirdly, the blue-shifted emission seen in C$^{18}$O agrees very well with the high-velocity large-scale emission seen in Fig. 3 of \citet{hasegawa95}. Therefore, if it were the case that the blue-shifted C$^{18}$O emission was created by VLA\,2, this source would also be responsible for the large-scale blue-shifted outflow seen by \citet{hasegawa95}. However, because this source has a much lower total luminosity compared to VLA\,3, which dominates the emission at all wavelengths, and as this would leave VLA\,3 without a corresponding blue-shifted outflow to the red-shifted one observed, we consider it more likely the blue-shifted emission is the inner part of the large-scale blue-shifted outflow of AFGL\,2591-VLA\,3.

Assuming that the observed blue-shifted C$^{18}$O emission is the inner part of the entrained large-scale blue-shifted outflow of AFGL\,2591-VLA\,3, the outflow properties were calculated, and are given in Table \ref{outflowtable}. The blue-shifted outflow mass was found assuming the gas was optically thin (derived in Appendix A),

\begin{figure*}[!htbp]
\begin{center}
\sidecaption
\includegraphics[width=12cm]{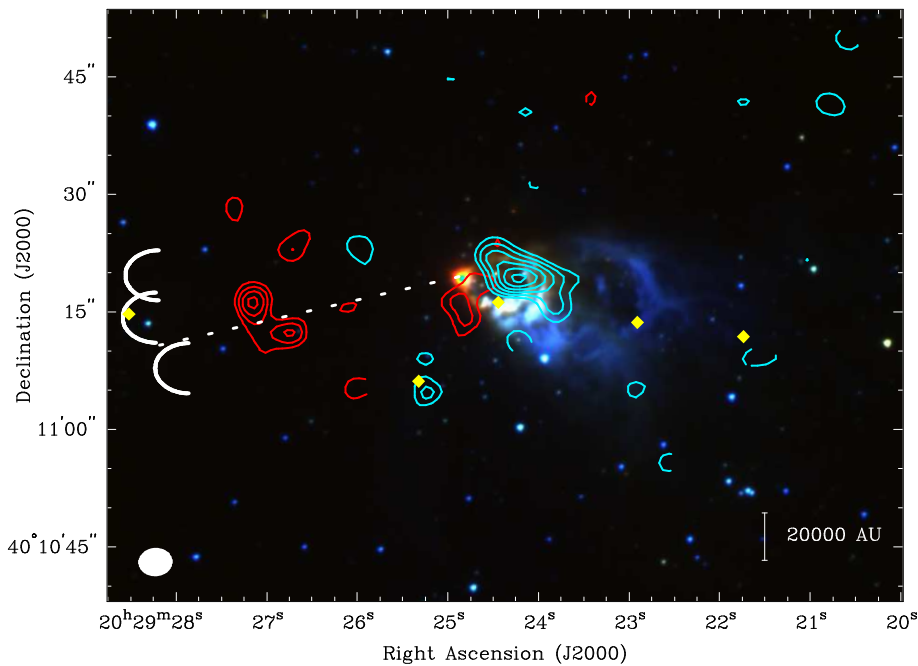}
 \caption[Gemini North three-colour JHK$'$ image of AFGL\,2591 overlaid with contours of C$^{18}$O emission integrated over red- and blue-shifted velocity ranges]{Gemini North three-colour JHK$'$ image of AFGL\,2591 overlaid with contours of C$^{18}$O emission integrated over -4.3 to -3.3\,km\,s$^{-1}$ (red) and -8.0 to -7.0\,km\,s$^{-1}$ (blue). The three arcs in the east of the image show the positions of bow shocks visible in the Gemini North image, which are too faint to be seen directly in the Figure. The dashed line shows the direction of the ionized jet observed at 3.6\,cm (P.A. 100$^{\circ}$), and yellow diamonds show roughly the peak positions of the H$_2$ knots observed in \citet{tamura92}. Red-shifted contours are -3,3,4,5,6 $\times$ 0.1\,Jy\,beam$^{-1}$\,km\,s$^{-1}$ and blue-shifted contours are -3,3,4,5,6,7 $\times$ 0.1\,Jy\,beam$^{-1}$\,km\,s$^{-1}$. The synthesised beam is shown in the lower left-hand corner (4.5$'' \times$ 3.6$''$, P.A. 93$^{\circ}$). Near-IR image stretch as in Fig.~\ref{cmGemfig}. \label{Gemini_compare}}
 \end{center}
\end{figure*}

\begin{equation}
M_{\rm{outflow}} = 2.75 \times 10^{-5} ~  \left[ \frac{^{12}\rm{CO}}{\rm{C}^{18}\rm{O}}\right] \frac{(T+0.882)}{e^{-5.27/T}}~d^2~ \int S_{\nu} ~d\upsilon~(\rm{M}_{\odot})
\label{massinflow}
\end{equation}

\noindent where $\left[ \frac{^{12}\rm{CO}}{\rm{C}^{18}\rm{O}}\right]$ is the abundance ratio of $^{12}\rm{CO}$ to $\rm{C}^{18}\rm{O}$, $T$ is the temperature of the gas assuming local thermodynamic equilibrium, $d_{kpc}$ is the distance to the source in kpc, and $\int S_{\nu} ~d\upsilon$ is the integral of the flux over the velocities of the blue-shifted emission. The flux in each channel was found by integrating the emission within an irregular aperture around only the blue-shifted outflow emission (i.e. not including emission from the resolved peak to the south of the central position of AFGL\,2591). The temperature was assumed to be 25\,K from the greybody fit to the far-IR and sub-mm fluxes in Section \ref{vla3}, and the abundance ratio  $\left[{^{12}\rm{CO}}/{\rm{C}^{18}\rm{O}}\right]$ was calculated using the results given in \citet{wilson94}, where $D_{_{GC}}$ is the galactocentric distance in kpc,

\begin{equation}
\left[ \frac{^{12}\rm{CO}}{\rm{C}^{18}\rm{O}}\right]= (58.8 \pm 11.8) D_{_{GC}} +(37.1 \pm 82.6),
\end{equation}

\noindent giving a value of $\left[ \frac{^{12}\rm{CO}}{\rm{C}^{18}\rm{O}}\right]=538$ for AFGL\,2591 for a galactocentric distance of 8.5\,kpc.

It should be noted that the shape of the $\rm{C}^{18}\rm{O}$ line profile shown in Fig. \ref{C18O_profile} may indicate that this line is partially optically thick, so that the estimates of the outflow properties depending on the derived mass in Table \ref{outflowtable} will be lower limits. In addition, as the $\rm{C}^{18}\rm{O}$ emission only traces the densest inner parts of the flow, it only represents a small fraction of the mass and momentum from the entire large-scale outflow.

To find the momentum $M_{\rm{outflow}}V$ and kinetic energy $\frac{1}{2}M_{\rm{outflow}}V^2$ in the flow, the final term of equation (\ref{massinflow}) was replaced by $\int S_{\nu} (\upsilon - \upsilon_{_{\rm LSR}})/\sin{i} ~d\upsilon$ or $\frac{1}{2}\int S_{\nu} ((\upsilon - \upsilon_{_{\rm LSR}})/\sin{i})^2 ~d\upsilon$ respectively, where $\upsilon$ is the velocity of each channel, $\upsilon_{_{\rm LSR}}=$-5.7\,km\,s$^{-1}$ is the rest velocity of the cloud, and $i$ is the inclination of the flow relative to the line of sight, taken to be $50^{\circ}$ from the results of the SED modelling presented in Section \ref{sedresults_afgl}.

\begin{table}
 \centering
  \caption{Derived properties of the inner part of the blue-shifted outflow traced by the C$^{18}$O emission}
  \begin{tabular}{@{}ll@{}}
  \hline
 Parameter     & Value \\
 \hline 
Velocity range (km\,s$^{-1}$) & -8 to -5.3 \\
Deprojected outflow length, $l$ (AU) & 61,000 \\
Outflow mass, $M_{\rm outflow}$ (M$_{\odot}$) & 133 \\
Outflow momentum, $M_{\rm outflow}V$ (M$_{\odot}$\,km\,s$^{-1}$)  & 164 \\
Outflow energy, $1/2\,M_{\rm outflow}V^2$ (M$_{\odot}$\,km$^{2}$\,s$^{-2}$)  & 178 \\
Outflow energy, $1/2\,M_{\rm outflow}V^2$ (erg)  & 3.54$\times$ 10$^{45}$ \\ 
Average velocity relative to $\upsilon_{_{LSR}}$, $\langle V \rangle$ (km\,s$^{-1}$) & -1.2 \\ 
Dynamical timescale, $t_{\rm dyn}$ (yr) & 2.4$\times$ 10$^5$ \\ 
\hline
\end{tabular}
\label{outflowtable}
\end{table}

The mass weighted average velocity of the flow was calculated as the momentum over the mass in the flow: $\langle V \rangle = M_{\rm{outflow}}V / M_{\rm{outflow}}$, and the projected length of the flow was measured as the distance from the central source of AFGL\,2591 to the furthest 3\,$\sigma$ contour from the source in the blue-shifted emission shown in Fig. \ref{Gemini_compare}, which was found to be $\sim$14.2$''$ or 47300\,AU at 3.33\,kpc. Therefore the actual length of the flow $l$ was calculated as $l=l_{\rm proj}/\sin{i}$. The dynamical timescale of the flow was determined using the equation $t_{\rm dyn}=l/\langle V \rangle$, however it should be noted that, as with any outflow, this value may not be an exact measure of the time taken to displace the gas, as the gas may have been set in motion at some distance from the central source. In addition, this assumes a constant outflow velocity, but the gas is observed to be accelerating away from the source. Nevertheless, this provides a characteristic timescale, which can be compared to other studies. Comparing to dynamical timescales found for large scale flows toward forming massive stars \cite[of order 10$^4$-10$^5$\,yr for $L>10^{4}$\,L$_{\odot}$,][]{beuther02},  the timescale for the inner part of the flow seen in C$^{18}$O is at the upper end of these values. Scaling the blue-shifted dynamical timescale of $6.3\times10^4$\,yr for the $^{12}$CO outflow observed by \citet{hasegawa95} to the new distance of 3.33\,kpc, we obtain a dynamical timescale of $2.1\times10^5$\,yr, similar to the value we find for the smaller-scale C$^{18}$O outflow.

\subsection{Cluster Properties}
\label{overall_properties}
The Section below deals with the overall properties of the forming cluster surrounding AFGL\,2591-VLA\,3. The maximum envelope radius for AFGL\,2591-VLA\,3, found by our modelling in Section \ref{sedresults_afgl}, is on parsec scales (180,000\,AU and 420,000\,AU or 0.87 and 2.0\,pc for the envelope with and without disk models respectively). This is also corroborated by the large-scale envelope observed in the $^{12}$CO and $^{13}$CO maps of \citet{van-der-wiel11}, as well as in the 80$''$ or 270,000\,AU diameter disk-like structure seen in the CS(1-0) maps of \citet{yamashita87}. This coherent parsec-scale structure, containing at least five forming stars (VLA\,1 through 5), therefore constitutes a cluster-forming clump.

The spectral types of the stars required to power VLA\,1, 2, 4 and 5 were calculated using the 3.6\,cm fluxes, by the same method as outlined in \citet{johnston09}, assuming they are optically thin H{\sc II} regions with one powering star, giving spectral types of B0.5, B1, B2 and B1 at 3.33\,kpc for VLA\,1, 2, 4 and 5 respectively. As VLA\,3 has a spectral index of between 0.4 and 0.5, its emission is not optically thin, and thus such an estimate would not be accurate. As the morphology is consistent with a compact core plus jet, it may be that some of its emission is not from an HC\,H{\sc II} region, but is instead created by shock-ionized gas. However, assuming photoionization dominates the emission from VLA\,3, the spectral type of the powering star of VLA\,3 was estimated to be O8-O9 under the assumption of a photoionized wind, using the mass loss rate of the jet found in Section \ref{outflow_jet_VLA3}: 0.77 - 1.0$\times10^{-5}\,\rm{M}_{\odot} \rm{yr}^{-1}$, the assumed wind velocity: 500\,km\,s$^{-1}$, and Table 4 of \citet{rodriguez83}. For the above spectral type estimates, we note that the fluxes in Table \ref{vlafluxtable} may only represent a fraction of the total flux at these wavelengths, due to the finite range in spatial scales which the observations probe, and therefore they provide lower limits.

Following \citet{sanna12}, we take the mass of the most luminous star VLA\,3 to be 38\,M$_{\odot}$. Assuming sources VLA\,1 to 5 are cluster members, the lowest luminosity powering star (VLA\,4 which is B2) corresponds to a ZAMS mass of $\sim$9\,M$_{\odot}$, found using the ZAMS luminosities of \citet{panagia73} and the theoretical HR diagram of \citet{maeder89}. Thus if this cluster has a fully populated IMF, extrapolating a Krupa IMF \citep{kroupa01} down to 1\,M$_{\odot}$ suggests there should be $\sim$100 stars in this cluster with mass above 1\,M$_{\odot}$.

The roughly circular shape of VLA\,1 (with a radius of $\sim$2$''$ or $\sim$6700\,AU at 3.33\,kpc) and its optically thin spectral index, supports the view that it is a UC\,H{\sc II} region excited by one or more massive stars. However, an exciting star for VLA\,1 is not seen in either the 2MASS or Gemini North near-IR images. Assuming a single star, the ZAMS B0.5 absolute magnitude given in \citet{panagia73} of $-2.8$, and a distance of 3.33\,kpc, the spectral type of the calculated exciting star of VLA\,1 corresponds to a 2MASS K$_s$ band magnitude of 11.1. The detection limit of the 2MASS point source catalog in K$_s$ band is 14.3 mag \citep{skrutskie06}, which indicates that there must be more than $3.2$\,mag of extinction at K$_s$ band for this source to be obscured, corresponding to $26$\,mag of extinction in the V band \citep{cardelli89}. These high extinctions, in addition to the indentation that VLA\,1 has made in the ambient cloud traced by near-IR emission, suggest that VLA\,1 lies within the same cloud core as AFGL\,2591-VLA\,3.

The dent in the cloud at the position of VLA\,1 provides support to the idea that this UC\,H{\sc II} region is at the same distance, and is expanding into the material illuminated by the central source of AFGL\,2591. Therefore it seems likely that the object at approximately 20$^{\rm h}$29$^{\rm m}$24$^{\rm s}$.6 +40$^{\circ}$11$'$16$''$.4 (J2000), suggested to be an edge-on disk by \citet{preibisch03}, is instead the brightened rim of the expanding UC\,H{\sc II} region, where it is interacting with the surrounding cloud. Hence, through its expansion, VLA\,1 may have been able to affect the formation of other members of the cluster.

The secondary peak in Fig. \ref{C18O_mom1} shows a velocity gradient across it in roughly the NW-SE direction. There are no sources at this position in any of the images, limiting the possibility that this emission is from an outflow of a nearby object. Instead the velocity of the emission appears to decrease radially away from VLA\,1, which is marked as a circle in Fig. \ref{C18O_mom1}. Indeed, Fig. \ref{cmGemfig} indicates that VLA\,1 is preferentially expanding towards the southeast, into possibly lower density gas. Hence the origin of this velocity gradient in Fig. \ref{C18O_mom1} may be due to the expansion of the gas around VLA\,1 towards the observer, with a projected velocity of $\sim$1\,km\,s$^{-1}$ relative to the ambient cloud. Further, it is interesting to note that the lack of C$^{18}$O emission at the position of VLA\,1 again suggests that VLA\,1 is at the same distance as AFGL\,2591-VLA\,3. The lack of C$^{18}$O emission towards VLA\,1 could be due to the fact it has swept-up the molecular cloud material around it, or the molecular gas has been dissociated by the high temperatures ($\sim10^4$\,K) in the H{\sc II} region.

\section{Conclusions}
\label{afgl:conclusions}

We have carried out a study of the emission from the luminous star-forming region AFGL\,2591 from near-IR through to cm wavelengths, to determine whether the properties of the source can be explained by the same processes thought to govern the formation of low-mass stars. 

In Section \ref{sedmodelling}, we described a Monte Carlo dust continuum radiative transfer dust code to model the continuum absorption, emission and scattering through two scaled-up azimuthally symmetric dust geometries, the first consisting of a rotationally flattened envelope with outflow cavities, and the second also including a flared disk. In Section \ref{sedresults_afgl} our results  show that in both cases, these models were able to reproduce the main features of the SED and 2MASS images of AFGL\,2591.

The observed 3.6\,cm images presented in Section \ref{obsresultcm} detect and more finely resolve the central powering source AFGL\,2591-VLA\,3, which has a compact core plus collimated jet morphology. The east component of the jet extends 4000\,AU from the central source, with an opening angle of $<10^{\circ}$ at this radius. This jet is also roughly parallel to the outflow seen at larger scales by \citet{hasegawa95}, confirming that VLA\,3 is the powering source of the outflow from AFGL\,2591. The multi-configuration 7\,mm image detects both VLA\,1 and VLA\,3. At this wavelength, VLA\,3 does not show a jet morphology, but instead presents compact ($<500$\,AU) emission, some of which ($<$0.57\,mJy of 2.9\,mJy) is estimated to be due to dust emission. Comparison with ancillary 0.3-0.4$''$ Gemini near-IR images show that VLA\,3 is coincident with the illuminating source of AFGL\,2591, at the apex of the observed near-IR reflection nebula. Further, VLA\,2 is coincident with the close binary first seen by \citet{preibisch03}, and VLA\,4 and a newly detected source VLA\,5, are coincident with near-IR sources. 

By fitting greybodies to the SEDs of VLA\,1 and VLA\,3 at wavelengths shorter than 1\,mm, the contribution of VLA\,1 to the total SED from AFGL\,2591 was determined to be $<30\%$ at wavelengths between 8 and 100$\mu$m, and to be negligible at other wavelengths. Therefore contamination by VLA\,1, or fainter sources in the region (e.g. VLA\,2), should not significantly change the emergent SED of the central powering object which dominates the luminosity from the region, AFGL\,2591-VLA\,3.

The nature of the emission from VLA3 was also investigated. The spectral index of VLA\,3 was calculated (between 3.6\,cm and 7\,mm) to be between 0.4 and 0.5, similar to that of an ionized wind. In Section \ref{outflow_jet_VLA3}, the 3.6\,cm emission was modelled as an ionized jet, from which the mass loss rate in the jet was determined to be in the range 0.77 - 1.0$\times10^{-5}$\,M$_{\odot}$yr$^{-1}$. In addition, it was found that the jet may have enough momentum to drive the larger-scale flow. However, assuming a shock efficiency of 10\%, the momentum rate of the jet was found to not be sufficient to ionize it via only shocks. Thus a significant portion of the 3.6\,cm emission is due to photoionization.

In Section \ref{obsresultmm}, the observed C$^{18}$O emission towards the source was found to be tracing the densest accelerated or entrained material. The blue-shifted emission, to the west of VLA\,3 and coincident with VLA\,2, shows an increasing velocity gradient to the west, which is more collimated at higher velocities, and is coincident with the inner part of the blue-shifted outflow observed by \citet{hasegawa95}. Thus although we cannot rule out that the that this emission is an outflow launched by VLA\,2, there is substantial evidence that the C$^{18}$O emission is tracing the densest parts of the entrained gas in the large-scale outflow from VLA\,3. In addition, the dense red-shifted gas traced by C$^{18}$O lies along the position angle of the ionized jet seen at 3.6\,cm, directly between the jet and bow shocks seen in the near-IR Gemini image. Therefore it is likely this emission also traces a small, dense, segment of the red-shifted flow from VLA\,3. 

The above results show that many of the properties of AFGL\,2591-VLA\,3 resemble those of low-mass protostars. For example, the geometry of AFGL\,2591 determined from modelling the SED and near-IR images of the source is consistent with that of a rotationally flattened envelope with and without a flared disk. Furthermore, the outflow of AFGL\,2591-VLA\,3 is comprised of both a collimated ionized jet, and a wide angle wind, a property seen in the outflow phenomena of many low-mass sources \citep[e.g.][]{arce07}. However, although the emission at 3.6\,cm from AFGL\,2591-VLA\,3 has the morphology of a jet, it is unlikely that it can be explained solely by shocks in a neutral wind or outflow. Therefore some part of the compact emission from the star may instead be provided by a hypercompact H{\sc II} region. Yet if this is the case, its presence has not disrupted the accretion or outflow processes of AFGL\,2591-VLA\,3. Thus, in this manner, the formation of the central dominant object VLA\,3 of AFGL\,2591 does not appear to be significantly different to that of low-mass protostars. However, it is important to note that this star is not forming in isolation, evidenced by the four other cm sources observed within the bounds of the envelope determined for AFGL\,2591-VLA\,3, with a radius found from our modelling to be on parsec scales. Therefore another way to view this picture is that AFGL\,2591-VLA\,3 is able to source its accreting material from the shared gas reservoir of a small cluster while still exhibiting the phenomena expected during the formation of low-mass stars.

\begin{acknowledgements}
We would like to thank the referee for providing insightful comments which improved this work. This publication makes use of data products from the Two Micron All Sky Survey, which is a joint project of the University of Massachusetts and the Infrared Processing and Analysis Center/California Institute of Technology, funded by the National Aeronautics and Space Administration and the National Science Foundation. 
This work makes use of observations obtained at the Gemini Observatory, which is operated by the 
Association of Universities for Research in Astronomy, Inc., under a cooperative agreement 
with the NSF on behalf of the Gemini partnership: the National Science Foundation (United 
States), the Science and Technology Facilities Council (United Kingdom), the 
National Research Council (Canada), CONICYT (Chile), the Australian Research Council (Australia), 
Minist\'{e}rio da Ci\^{e}ncia e Tecnologia (Brazil) 
and Ministerio de Ciencia, Tecnolog\'{i}a e Innovaci\'{o}n Productiva (Argentina).
We also used APLpy, an open-source plotting package for Python hosted at http://aplpy.github.com.
\end{acknowledgements}

\begin{appendix}
\label{appendix}
\section{Mass determination from C$^{18}$O emission}

Starting from equation A1 from the appendix in \citet{scoville86}:

\begin{equation}
N_{total}= \frac{3k}{8 \pi^3 B_e {\mu}^2} \frac{e^{hB_e J_l(J_l+1)/kT}}{(J_l+1)} \frac{T+hB_e/3k}{\left(1- \exp{\left(-h\nu/kT\right)} \right)} ~ \int_0^{\infty} \tau_{\upsilon} ~d\upsilon
\label{scovilleA1}
\end{equation}

\noindent where $k$ is the Boltzmann constant, $B_e$ is the rotational constant for the molecule, $\mu$ is the dipole moment of the molecule in e.s.u. (1 Debye=$10^{-18}$ e.s.u.), $h$ is the planck constant, $J_l$ is the rotational quantum number for the lower energy level, $T$ is the excitation temperature assumed to characterise the populations in all of the energy levels of the gas, $\nu$ is the frequency of the emission and $\tau_{\upsilon}$ is the optical depth over the Doppler line profile as a function of velocity $\upsilon$. 

The solution to the equation of radiative transfer (assuming background terms are negligible) is

\begin{equation}
T_b = \frac{h\nu/k}{e^{h\nu/kT} -1} (1 - e^{-\tau_{\upsilon}}).
\label{solution}
\end{equation}

Multiplying equation \ref{scovilleA1} by the ratio of the LHS to RHS of equation (\ref{solution}) and simplifying the terms containing $e^{-h\nu/kT}$ gives

\begin{align}
\begin{split}
{N}_{total} = & \frac{ {N}_{total} ~ T_b (e^{h\nu/kT} -1)} {h\nu/k (1 - e^{-\tau_{\upsilon}}) }\\
= & \frac{3k^2}{8 \pi^3 B_e {\mu}^2 h \nu} \frac{e^{hB_e J_l(J_l+1)/kT}}{(J_l+1)} \frac{T+hB_e/3k}{e^{-h\nu/kT}} \int_0^{\infty} \frac{T_b ~ \tau_{\upsilon}}{(1-e^{-\tau_{\upsilon}})} ~d\upsilon.
\end{split}
\label{nbar}
\end{align}

\noindent The total mass of gas in the source is given by:

\begin{equation}
M_{H_{2}}= {N}_{total} \left[ \frac{H_2}{^{12}\rm{CO}}\right] \left[ \frac{^{12}\rm{CO}}{\rm{C}^{18}\rm{O}}\right] \mu_{_G} m_{_H{_{_2}}} \frac{\pi \theta^2}{4} d^2
\label{massh2}
\end{equation}
 
\noindent where $\left[ \frac{H_2}{^{12}\rm{CO}}\right]$ is the H${_{_2}}$ to $^{12}$CO abundance ratio, $\left[ \frac{^{12}\rm{CO}}{\rm{C}^{18}\rm{O}}\right]$ is the $^{12}$CO to C$^{18}$O abundance ratio, $\mu_{_G}$ is the mean atomic weight of the gas, $m_{_H{_{_2}}}$ is the molecular mass of $H{_{_2}}$, $\theta$ is the angular diameter of a uniform disk source, and $d$ is the distance to the source.

Substituting ${N}_{total}$ (equation \ref{nbar}) into equation (\ref{massh2}) we find:

\begin{equation}
\begin{split}
M_{H_{2}}= \frac{3k^2}{8 \pi^3 B_e {\mu}^2 h \nu} & \frac{e^{hB_e J_l(J_l+1)/kT}}{(J_l+1)} \frac{T+hB_e/3k}{e^{-h\nu/kT}} \int_0^{\infty} \frac{T_b ~ \tau_{\upsilon}}{(1-e^{-\tau_{\upsilon}})}  ~d\upsilon \\
& \left[ \frac{H_2}{^{12}\rm{CO}}\right]  \left[ \frac{^{12}\rm{CO}}{\rm{C}^{18}\rm{O}}\right]\mu_{_G} m_{_H{_{_2}}} \frac{\pi \theta^2}{4} d^2
\label{massh2_2}
\end{split}
\end{equation} 

Constants used as well as several values specific to the {(1-0)} transition of C$^{18}$O are listed below, expressed in c.g.s. units. Values for $B_e$ are taken from \citet{rosenblum58}, and values for $\mu$ are taken from the JPL line catalog \footnote{http://spec.jpl.nasa.gov/ftp/pub/catalog/catdir.html}.
\begin{align*}
& J_l = 0 \\
& \nu = 1.09782182 \times 10^{11}~ \rm{Hz} \\
& B_e = 5.5135449 \times 10^{10}~ \rm{cycles/s}\\
& \mu = 0.11079 ~\rm{Debye} = 0.11079 \times 10^{-18} \rm{e.s.u} \\ 
& \left[ \frac{H_2}{^{12}\rm{CO}}\right] = 10^4 \\
& \mu_g = 1.36 \\
& m_{H_2} = 3.34524316 \times 10^{-24}\,\rm{g} \\
\end{align*}
 
Entering these into equation (\ref{massh2_2}) we obtain:
 
\begin{equation}
\begin{split}
M_{H_{2}} = & 1.88 \times 10^{-7} ~ \left[ \frac{^{12}\rm{CO}}{\rm{C}^{18}\rm{O}}\right] \frac{(T+0.882)}{e^{-5.27/T}} \\
& \theta_{arcsec}^2 d_{kpc}^2~ \int_0^{\infty} \frac{T_b ~ \tau_{\upsilon}}{(1-e^{-\tau_{\upsilon}})} ~d\upsilon \rm{~~M}_{\odot}
\end{split}
\end{equation}
  
Similarly, the mass of gas may be obtained from the line flux density $S_{\nu}$, integrated over a map, 

\begin{align}
T_b &= \frac{c^2}{2k}\frac{1}{\nu^2} S_{\nu},
\end{align}

\noindent we find an alternative expression for the mass of the gas:
 
\begin{equation}
\begin{split}
M_{H_{2}} = &~2.75 \times 10^{-5} ~ \left[ \frac{^{12}\rm{CO}}{\rm{C}^{18}\rm{O}}\right]  \frac{(T+0.882)}{e^{-5.27/T}} \\
& d_{kpc}^2~ \int_0^{\infty} \frac{S_{\nu} ~ \tau_{\upsilon}}{(1-e^{-\tau_{\upsilon}})} ~d\upsilon \rm{~~M}_{\odot}
\end{split}
\label{finalwithtau}
\end{equation}
 
\noindent where $S_{\nu}$ is measured in Jy. 

If the assumption is made that the emitting gas is optically thin ($\tau\ll1$) over the line, equation (\ref{finalwithtau}) simplifies to

\begin{align}
M_{H_{2}} = 2.75 \times 10^{-5} ~  \left[ \frac{^{12}\rm{CO}}{\rm{C}^{18}\rm{O}}\right]  \frac{(T+0.882)}{e^{-5.27/T}} d_{kpc}^2~ \int_0^{\infty} S_{\nu} ~d\upsilon \rm{~~M}_{\odot}.
\end{align}

The abundance ratio  $\left[{^{12}\rm{CO}}/{\rm{C}^{18}\rm{O}}\right]$ can be calculated using the results given in \citet{wilson94}, where $D_{_{GC}}$ is the galactocentric distance in kpc:

\begin{equation}
\left[ \frac{^{12}\rm{CO}}{\rm{C}^{18}\rm{O}}\right]= (58.8 \pm 11.8) D_{_{GC}} +(37.1 \pm 82.6).
\end{equation}

\end{appendix}

\bibliography{} 

\end{document}